%% file: iphas-dr2-rev2-arxiv.tex
\documentclass[a4paper,useAMS,usenatbib]{mn2e}
\pdfpagewidth=\paperwidth
\pdfpageheight=\paperheight
\usepackage{amssymb,amsmath}
\usepackage[pdftex]{graphicx}
\usepackage{longtable} 
\usepackage{float}
\usepackage{booktabs}
\usepackage{aas_macros}
\usepackage{caption}
\usepackage{subcaption}

\def\ha{\mbox{H$\rm \alpha$}}
\def\micron{\mbox{$\mu$m}}

\title[IPHAS Data Release 2]{The Second Data Release 
of the INT Photometric H$\alpha$ Survey 
of the Northern Galactic Plane (IPHAS DR2)}
		
\author[G. Barentsen
et al.]{Geert Barentsen$^{1}$\thanks{E-mail:geert@barentsen.be},
H. J. Farnhill$^1$,
J. E. Drew$^1$,
E. A. Gonz$\acute{\rm{a}}$lez-Solares$^2$, \newauthor
R. Greimel$^3$, 
M. J. Irwin$^2$,
B. Miszalski$^{4}$,
C. Ruhland$^1$,
P. Groot$^5$,
A. Mampaso$^{6,7}$, \newauthor
S. E. Sale$^8$, 
A. A. Henden$^9$,
A. Aungwerojwit$^{10}$,
M. J. Barlow$^{11}$,
P. J. Carter$^{12}$, \newauthor
R. L. M. Corradi$^{6,7}$, 
J. J. Drake$^{13}$, 
J. Eisl\"offel$^{14}$, 
J. Fabregat$^{15}$, 
B. T. G\"ansicke$^{12}$, \newauthor
N. P. Gentile Fusillo$^{12}$, 
S. Greiss$^{12}$, 
A. S. Hales$^{16}$, 
S. Hodgkin$^2$,
L. Huckvale$^{17}$, \newauthor
J. Irwin$^{13}$,
R. King$^{18}$,
C. Knigge$^{19}$, 
T. Kupfer$^{5}$,
E. Lagadec$^{20}$,
D. J. Lennon$^{21}$, \newauthor
J. R. Lewis$^{2}$,
M. Mohr-Smith$^{1}$,
R. A. H. Morris$^{22}$, 
T. Naylor$^{18}$, 
Q. A. Parker$^{23,24,25}$,  \newauthor
S. Phillipps$^{21}$, 
S. Pyrzas$^{26}$,
R. Raddi$^{12}$,
G. H. A. Roelofs$^{13}$, 
P. Rodr\'\i guez-Gil$^{6,7}$, \newauthor
L. Sabin$^{27}$, 
S. Scaringi$^{28,29}$,
D. Steeghs$^{12}$,
J. Suso$^{14}$,
R. Tata$^{6,7}$,
Y. C. Unruh$^{30}$, \newauthor
J. van Roestel$^5$, 
K. Viironen$^{31}$, 
J. S. Vink$^{32}$, 
N. A. Walton$^{2}$, 
N. J. Wright$^{1}$, \newauthor
A. A. Zijlstra$^{17}$.
\\
$^{1}$School of Physics, Astronomy \& Mathematics, University of Hertfordshire, College Lane, Hatfield, Hertfordshire, AL10 9AB, U.K.\\
$^{2}$Institute of Astronomy, University of Cambridge, Madingley Road, Cambridge, CB3 OHA, U.K.\\
$^{3}$IGAM, Institute of Physics, University of Graz, Universit\"atsplatz 5, 8010 Graz, Austria.\\
$^{4}$South African Astronomical Observatory \& SALT Foundation, P.O. Box 9, Observatory, 7935 Cape Town, South Africa.\\
$^{5}$Afdeling Sterrenkunde, Radboud Universiteit Nijmegen, Faculteit NWI, Postbus 9010, 6500 GL Nijmegen, The Netherlands.\\
$^{6}$Instituto de Astrof\'\i sica de Canarias, V\'\i a L\'actea, s/n, La Laguna, E-38205, Santa Cruz de Tenerife, Spain.\\
$^{7}$Departamento de Astrof\'\i sica, Universidad de La Laguna, La Laguna, E-38204, Santa Cruz de Tenerife, Spain.\\
$^{8}$Rudolf Peierls Centre for Theoretical Physics, Keble Road, Oxford, OX1 3NP, U.K.\\
$^{9}$AAVSO, 49 Bay State Road, Cambridge, MA 02138, USA.\\
$^{10}$Department of Physics, Faculty of Science, Naresuan University, Phitsanulok 65000, Thailand.\\
$^{11}$University College London, Department of Physics \& Astronomy, 
Gower Street, London WC1E 6BT, U.K.\\
$^{12}$Department of Physics, University of Warwick, Gibbet Hill Road, Coventry, CV4 7AL, U.K.\\
$^{13}$Harvard-Smithsonian Center for Astrophysics, 60 Garden Street, 
Cambridge, MA 02138, U.S.A. \\
$^{14-32}$ (affiliations 14 through 32 are given in the source file)
}

\begin{document}
\date{Current draft typeset \today}
\pagerange{\pageref{firstpage}--\pageref{lastpage}} \pubyear{2014}
\maketitle

\label{firstpage}

\begin{abstract}
The INT/WFC Photometric H$\alpha$ Survey 
of the Northern Galactic Plane (IPHAS)
is a 1800~deg$^2$ imaging survey
covering Galactic latitudes $|b| < 5\degr$
and longitudes $\ell=30\degr$ to 215\degr\ 
in the $r$, $i$ and \ha\ filters 
using the Wide Field Camera (WFC) 
on the 2.5-metre Isaac Newton Telescope (INT) in La Palma.
We present the first quality-controlled and
globally-calibrated source catalogue
derived from the survey,
providing single-epoch photometry
for 219 million unique sources
across 92 per cent of the footprint.
The observations were carried out between 2003 and 2012
at a median seeing of 1.1~arcsec (sampled at 0.33~arcsec/pixel)
and to a mean $5\sigma$-depth of 
21.2~($r$), 20.0~($i$) and 20.3~($\ha$)
in the Vega magnitude system.
We explain the data reduction 
and quality control procedures,
describe and test the global re-calibration,
and detail the construction of the new catalogue.
We show that the new calibration is accurate to
0.03~mag (rms) 
and recommend a series of quality criteria
to select accurate data
from the catalogue.
Finally, we demonstrate the ability of the 
catalogue's unique
$(r-\ha,\ r-i)$ diagram to
(i) characterise stellar populations and extinction regimes
towards different Galactic sightlines
and (ii) select and quantify \ha\ emission-line objects.
IPHAS is the first survey to offer comprehensive CCD photometry
of point sources across the Galactic Plane at visible wavelengths,
providing the much-needed counterpart
to recent infrared surveys.
\end{abstract}


\section{Introduction}
The INT/WFC Photometric H$\alpha$ Survey
of the Northern Galactic Plane \citep[IPHAS;][]{Drew2005}
is providing new insights into the contents and structure of the disk of the Milky Way.
This large-scale programme of observation
-- spanning a decade so far 
and using more than 300 nights in competitive open time 
at the Isaac Newton Telescope (INT) in La Palma --
aims to provide the digital update 
to the photographic northern H$\alpha$ surveys 
of the mid-20th century \citep[see][]{Kohoutek1999}. 
By increasing the sensitivity 
with respect to these preceding surveys 
by a factor $\sim$1000 (7~magnitudes), 
IPHAS can expand
the limited bright samples of Galactic emission line objects 
previously available into larger, deeper,
and more statistically-robust samples
that will better inform our understanding 
of the early and late stages of stellar evolution.
Since the publication of the Initial Data Release \citep[IDR;][]{Gonzalez-Solares2008},
these aims have begun to be realised through a
range of published studies including: 
a preliminary catalogue of candidate emission-line objects \citep{Witham2008};
discoveries of new symbiotic stars \citep{Corradi2008, Corradi2010, Corradi2011, Rodriguez2014}; 
new cataclysmic variables \citep{Witham2007,Wesson2008,Aungwerojwit2012}; 
new groups of young stellar objects
\citep{Vink2008,Barentsen2011a,Wright2012};
new classical Be stars \citep{Raddi2013};
along with discoveries of new supernova remnants \citep{Sabin2013}
and new and remarkable planetary nebulae 
\citep{Mampaso2006, Viironen2009a, Viironen2009b, Sabin2010, Corradi2011, Viironen2011,Sabin2014}.

Over the years it has become apparent that the legacy of IPHAS 
will reach beyond these traditional \ha\ applications 
of identifying emission-line stars and nebulae. 
Through the provision of $r$ and $i$ broadband photometry 
alongside narrowband \ha\ data,
IPHAS has created the opportunity 
to study Galactic Plane populations 
in a new way.
For example, the survey's unique $(r-\ha,\ r-i)$ colour-colour
diagram has been shown to provide simultaneous constraints 
on intrinsic stellar colour and interstellar extinction \citep{Drew2008}. 
This has opened the door 
to a wide range of Galactic science applications, 
including the mapping of extinction across the Plane in three dimensions
and the probabilistic inference of stellar properties
\citep{Sale2009, Sale2010, Giammanco2011, Sale2012, Barentsen2013, Sale2014}. 
In effect, the availability of narrowband \ha\
alongside $r$ and $i$ magnitudes
provides coarse spectral information for huge samples of stars 
which are otherwise too faint or numerous 
to be targeted by spectroscopic surveys (cf. the use of
Stromgren $uvbyH\beta$ photometry at blue wavelengths).
For such science applications to succeed, however, 
it is vital that the imaging data are transformed 
into a homogeneously-calibrated photometric catalogue, 
in which quality problems 
and duplicate detections are flagged. 

When the previous release -- the IDR -- was created in late 2007,
just over half of the survey footprint was covered
and the data were insufficiently complete 
to support a homogeneously calibrated source catalogue.
The goal of this paper is to present the next release, 
which supersedes the IDR by including a global re-calibration
and by taking the coverage up to 92 per cent of the survey area.
In this work we
(i) explain the data reduction 
and quality control procedures that were applied,
(ii) describe and test the new photometric calibration, and 
(iii) detail the construction of the source catalogue
and demonstrate its use.

In \S\ref{sec:observations} we start by recapitulating the key points
of the survey observing strategy.
In \S\ref{sec:reduction} we describe the data reduction
and quality control procedures.
In \S\ref{sec:calibration} we explain the global re-calibration, in which
we draw upon the AAVSO Photometric All-Sky Survey (APASS)
and test our results against the Sloan Digital Sky Survey (SDSS).
In \S\ref{sec:catalogue} we explain how the source catalogue was compiled.
In \S\ref{sec:discussion} we discuss the properties of the catalogue
and in \S\ref{sec:demonstration} we demonstrate
the scientific exploitation of the colour/magnitude diagrams.
Finally, in \S\ref{sec:dataaccess} we discuss access
to the catalogue and an online library of reduced images.
The paper ends with conclusions in \S\ref{sec:conclusions} where we also outline
our future ambitions.

\section{Observations}
\label{sec:observations}

\begin{table}
    \caption{Key properties of IPHAS DR2.}
    \label{tbl:survey}
    \begin{center}
        \begin{tabular}{ll}
        \toprule
        Property & Value \\
        \midrule
        Telescope & 2.5m Isaac Newton Telescope \\
        Instrument & Wide Field Camera (WFC) \\
        Detectors & Four 2048 $\times$ 4100 pixel CCD's \\
        Pixel scale & 0.33 arcsec/pixel \\        
        Filters & $r$, $i$, \ha \\
        Filter properties & See Table~2 \\
		Magnitude system & Vega \\
        Exposure times & 30~s ($r$), 10~s ($i$), 120~s (\ha) \\
		Saturation limit & 13 ($r$), 12 ($i$), 12.5 (\ha) \\
        Detection limit (5$\sigma$, mean) & 21.2 ($r$), 20.0 ($i$), 20.3 (\ha) \\
        PSF FWHM (median) & 1.1\arcsec\ ($r$), 1.0\arcsec\ ($i$), 1.1\arcsec\ (\ha) \\
        Survey area & $\sim 1860$~deg$^2$ \\
        Footprint boundaries & $ -5\degr < b < +5\degr,\ 29\degr < \ell < 215\degr$ \\
        Observing period & August 2003 - November 2012 \\
        Website & www.iphas.org \\
        \bottomrule
        \end{tabular}
	\end{center}
\end{table}

The detailed properties of the IPHAS observing programme 
have been presented before 
by \citet{Drew2005} and \citet{Gonzalez-Solares2008}. 
To set the stage for this release,
we recap some key points in this section 
(see Table~1 for a quick-reference overview).
IPHAS is an imaging survey of the Galactic Plane north of the celestial equator, 
from which photometry in Sloan $r$ and $i$ 
is extracted along with narrowband \ha. 
It is carried out using the Wide Field Camera (WFC) 
on the 2.5-metre INT in La Palma. 
It is the first digital survey to offer comprehensive optical CCD photometry
of point sources in the Galactic Plane;
the footprint spans a box 
of roughly 180 by 10 degrees, 
covering Galactic latitudes $-5\degr < b < +5\degr$ 
and longitudes $30\degr < \ell < 215\degr$.

\begin{figure}
    \includegraphics[width=\linewidth]{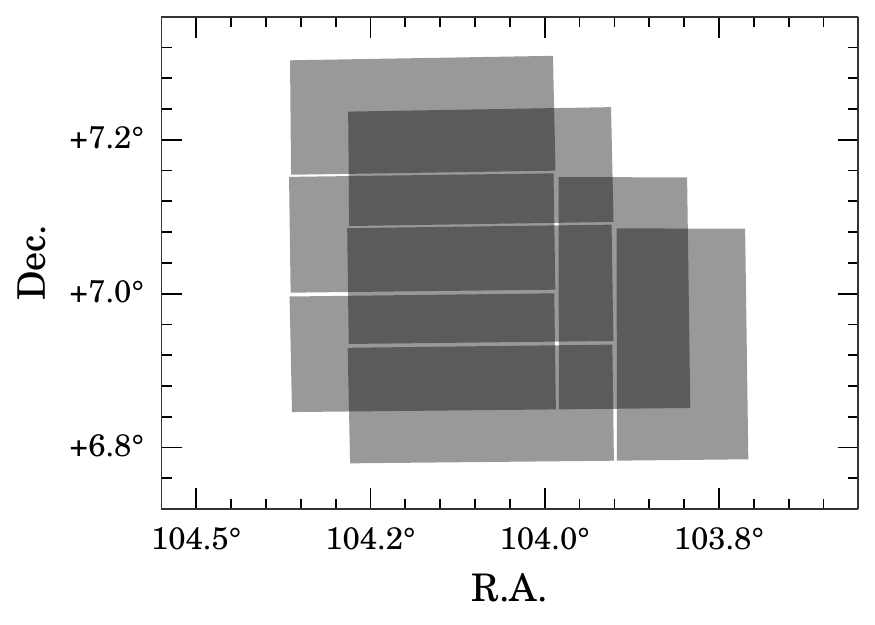} 
    \caption{
    Example footprint of a pointing and its offset position.
    Each field in the survey is accompanied by an offset field
    at 5\arcmin\ West and South
	to deal with gaps between the CCDs.
	This unit of observation is called a \emph{field pair},
	which is observed in all three filters within a span of 10 minutes.
	The WFC is a mosaic of four CCDs,
	and hence a field pair is composed of eight CCD frames.
	In this example, each CCD is plotted as a semi-transparent grey rectangle
	to highlight the overlap regions.
	Note that the L-shaped arrangement of the CCD mosaic allows
	nearly the entire field to be captured using just two pointings,
	apart from two $\sim10\arcsec$-wide squares
	which are located where inter-CCD gaps overlap.	
	}
    \label{fig:fieldpair}
\end{figure}

The WFC is a mosaic of 4 CCDs 
that captures a sky area of close to 0.29~deg$^2$ at a pixel scale of
0.33 arcsec/pixel.
To cover the Northern Plane with some overlap,
the survey area was divided into 7,635 telescope pointings.
Each of these pointings is accompanied by an offset position
displaced by $+$5 arcmin in Dec 
and $+$5 arcmin in RA,
to deal with inter-CCD gaps, detector imperfections,
and to enable quality checks.
An example footprint of a pointing
and its offset position is shown in Fig.~\ref{fig:fieldpair}.
Hence, the basic unit of observation
amounts to $2 \times 3$ exposures, 
in which each of the 3 survey filters is exposed at 2 offset sky positions 
within, typically, an elapsed time of 10 minutes.
We shall refer to the unit of 3 exposures at the same position 
as a \emph{field},
and the combination of two fields at a small offset as a \emph{field pair}.
Altogether, the survey contains 15,270 fields
grouped into 7,635 field pairs.
To achieve the desired survey depth
of 20th magnitude or fainter in each filter, 
the exposure times were set at 120~s (\ha), 
30~s ($r$) and 10~s ($i$)
in the vast majority of the survey observations.\footnote{The $r$-band exposure time was 10s instead of 30s in the first months of data taking. Since October 2010, the $i$-band exposure time 
has been increased from 10s to 20s to by-pass a sporadic exposure timing bug that affects the WFC.}

Data-taking began in the second half of 2003, 
and every field had been observed at least once by the end of 2008.
At that time only 76 per cent of the field pairs 
satisfied our minimum quality criteria, however.
The problems affecting the 24 per cent falling below survey standard were,
most commonly: variable cloud cover;
poor seeing; technical faults
(the quality criteria will be detailed in the next section).
Since then, a programme of repeat observations has been in place 
to improve data quality. 
As a result, 92 per cent of the survey footprint
now benefits from quality-approved data.
The most recent observations included in this release
were obtained in November 2012.

\begin{figure*}
        \includegraphics[width=1\linewidth]{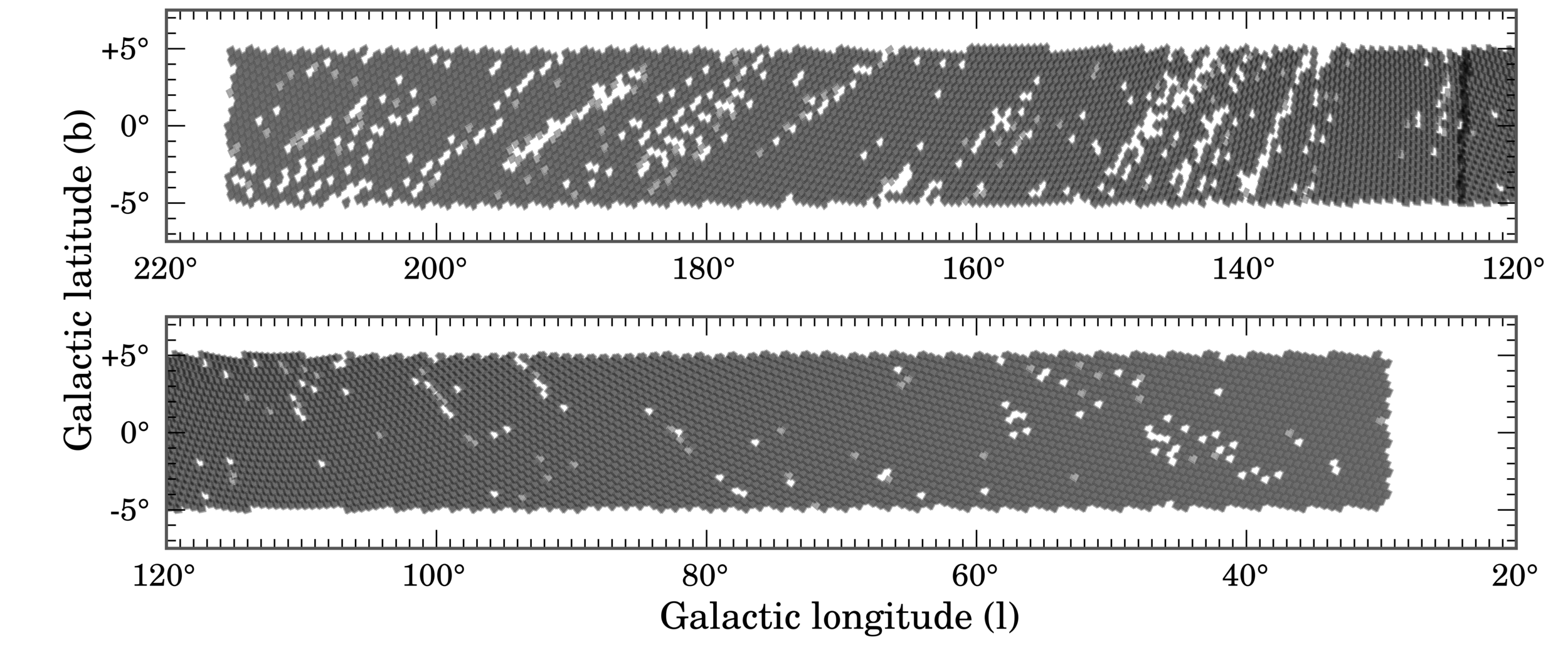}
        \caption{Survey area showing the footprints
        of all the quality-approved IPHAS fields
        which have been included in this data release.
        The area covered by each field has been coloured black
        with a semi-transparent opacity of 20 per cent,
        such that regions where fields overlap are darker.
        The IPHAS strategy is to observe each field twice
        with a small offset,
        and hence the vast majority of the area 
        is covered twice (dominant grey colour).
        There are small overlaps between all the neighbouring fields
        which can be seen as a honeycomb pattern
        of dark grey lines across the survey area.
        Regions with incomplete data are apparent as white gaps (no data) 
        or in light grey (indicating that one offset is missing).
        The dark vertical strip near $\ell \simeq$125\degr\ 
        is an arbitrary consequence of the tiling pattern,
        which was populated starting from 0h in Right Ascension.}
        \label{fig:footprint}
\end{figure*}

Fig.~\ref{fig:footprint} shows the footprint
of the quality-approved observations included in this work. 
The fields which remain missing 
-- covering 8 per cent of the survey area --
are predominantly located towards the Galactic anti-centre 
at $\ell > 120\degr$.
Fields at these longitudes are mainly accessed from La Palma 
in the months of November-December,
which is when the La Palma weather and seeing conditions are often poor,
forcing many (unsuccessful) repeat observations.
To enable the survey to be brought to completion, 
a decision was made recently to limit repeats in this area 
to individual fields requiring replacement,
i.e. fresh observations in all 3 filters may only be obtained 
at one of the two offset positions.  
The catalogue is structured such that it is clear 
where contemporaneous observations of both halves of a field pair
are available.

\section{Data reduction and quality control}
\label{sec:reduction}

\subsection{Initial pipeline processing}

All raw IPHAS data were transferred to the Cambridge Astronomical Survey Unit 
(CASU) for initial processing and archiving.  The procedures used by CASU 
were originally devised for the INT Wide Field imaging Survey 
\citep[WFS;][]{McMahon2001,Irwin2005}, which was a 200 deg$^2$ 
extragalactic survey programme carried out between 1998 and 2003.
Because IPHAS uses the same telescope and camera combination, we have been 
able to benefit from the existing WFS pipeline. A description of the 
processing steps can be found in \citet{Irwin2001}. Its application to IPHAS 
has previously been described
by \citet{Drew2005} and \citet{Gonzalez-Solares2008},
and some elements of the the source code are available 
on line\footnote{http://casu.ast.cam.ac.uk/surveys-projects/software-release}. 
In brief, the imaging processing part of the pipeline takes care of bias 
subtraction, the linearity correction, flat-fielding with internal gain
correction, and de-fringing for the i-band.

Object detection and parametrisation is then carried out
using the standard methods developed by CASU,
which can be summarised in four steps
\citep[a discussion on each of these steps
and related points can be found in][]{Irwin1985,Irwin1997}:
\begin{enumerate}
\item The local sky background is estimated by first computing an iteratively
sigma-clipped median intensity on a grid of 64$\times$64 pixel bins across 
the image from each detector.  This is usually robust against contaminating 
sources corrupting the background level. The resulting array of background
values is then filtered to further reduce the effect of large objects on
the local background level. Bilinear interpolation is then used to obtain an 
estimate of the background level at the orginal image pixel scale.
\item To improve faint object detection, each image is smoothed with a matched
detection filter which is used in conjunction with the unsmoothed image
for object detection and parameterisation.
\item Objects, or blends of objects, are detected
by identifying groups of 4 or more neighbouring contiguous pixels
in which the intensity exceeds the background level by at least $1.25\sigma$
on the matched filtered image. Objects are deblended using a sequence of
successively higher detection levels.
\item The objects are parametrised using the unsmoothed image at these pixel 
locations : positions are obtained based on an intensity-weighted isophotal 
centre-of-gravity of each object; whilst photometry is derived by measuring 
the intensity in a series of soft-edged circular apertures covering a range 
of diameters (1\farcs 2, 2\farcs 3, 3\farcs 3, 4\farcs 6, 6\farcs 6 and 
9\farcs 2).  Objects are also classified morphologically
-- stellar, extended or noise -- based on their intensity as a function of 
aperture size and on their intensity-weighted second moments, where the latter
are used to derive an estimate of object ellipticity.
\item The parametrisation of overlapping objects is refined by simultaneous 
fitting of soft-edged apertures to each blend,
effectively carrying out ``top hat'' profile fitting where it is necessary.
We note that the parametrisation of these blended objects is naturally
less reliable than that of single, unconfused sources,
which is why they are flagged in the catalogue
(to be explained in \S\ref{sec:catalogue}).
\end{enumerate}

Having carried out object detection,
the astrometric calibration is then determined.
The solution starts with a rough 
World Coordinate System (WCS)
based on the known telescope and camera geometry,
using a Zenithal Polynomial projection
\citep[ZPN;][]{Calabretta2002}
to model the (fixed) field distortion of the camera.
The parameters of this solution
are then progressively refined by fitting against the
Two-Micron All Sky Survey \citep[2MASS; ][]{Skrutskie2006},
albeit without correcting for the $\sim$10 years of proper motion
between the IPHAS and 2MASS epochs.
The resulting fit has previously been shown
to deliver results which are internally consistent
to better than 0\farcs 1 across the detector array
\citep{Gonzalez-Solares2008}.

\begin{figure}
    \includegraphics[width=\linewidth]{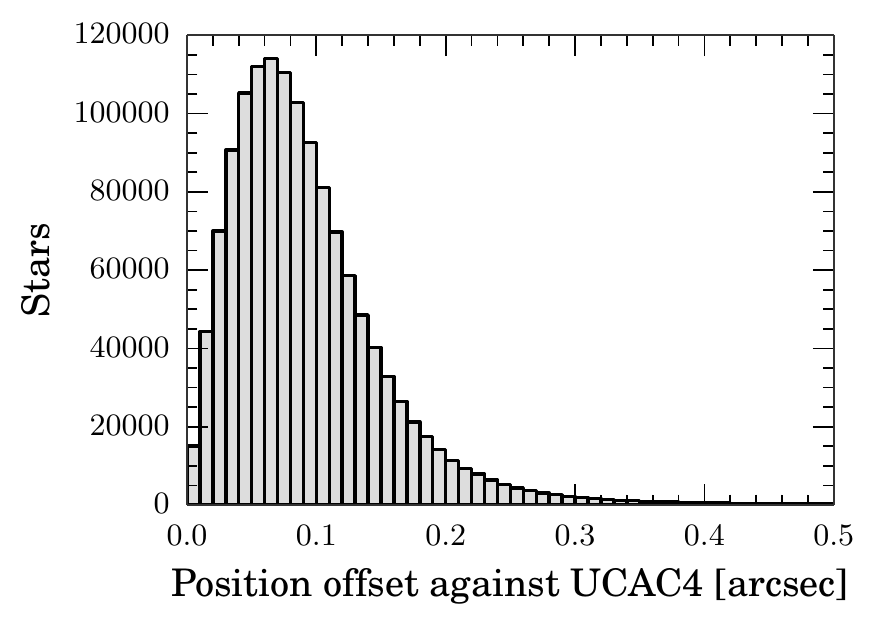} 
    \caption{Distribution of the astrometric residuals of stars
    		which appear both in IPHAS DR2 and UCAC4
    		within a cross-matching distance of 1\arcsec.
            The residuals were computed for 
            the 1.3 million stars in the IPHAS catalogue
            which are not blended, not saturated,
            and fall in the magnitude range $13 < r < 15$.
            The mean and standard deviation of this distribution 
            equals $94\pm65$~mas.}
    \label{fig:astrometry}
\end{figure}

An external validation of the astrometry 
has been carried out by comparing our positions against the
United States Naval Observatory CCD Astrograph Catalog 
\citep[UCAC4;][]{UCAC4}.
Fig.~\ref{fig:astrometry} shows the distribution
of the astrometric offsets
computed for 1.3 million stars
in the magnitude range $13 < r < 15$,
which is the range where both surveys overlap
and where the formal mean error of UCAC4
is better than 50~mas \citep{UCAC3}.
We find the mean difference in position 
between IPHAS and UCAC4 to be 94~mas,
which is satisfactory for our purposes.
These residuals are in part due to
the motion of the Earth through our Galaxy,
which we did not account for.

At the time of preparing DR2,
the pipeline had processed
74,195 single-band IPHAS exposures 
in which a total of 1.9~billion \emph{candidate} detections
were made.
This total inevitably includes spurious objects, artefacts and
duplicate detections;
in \S\ref{sec:catalogue} we will explain
how these have been removed or flagged in the final catalogue.

The entire data set -- comprising 2.5~terabyte of FITS files --
was then transferred to the University of Hertfordshire
for the purpose of transforming the raw
detection tables into a source catalogue which
(i) is quality-controlled,
(ii) is homogeneously calibrated, and 
(iii) contains user-friendly columns and warning flags.
It is these post-processing steps which distinguish this release
from the IDR, which (i) enforced less stringent quality limits,
(ii) did not offer a global calibration,
and (iii) provided a catalogue in which duplicate detections
of unique sources were not flagged.
These improvements are explained next.

\subsection{Quality control}
\label{sec:qc}

\begin{figure}
    \begin{minipage}[b]{\linewidth}
        \includegraphics[width=\textwidth]{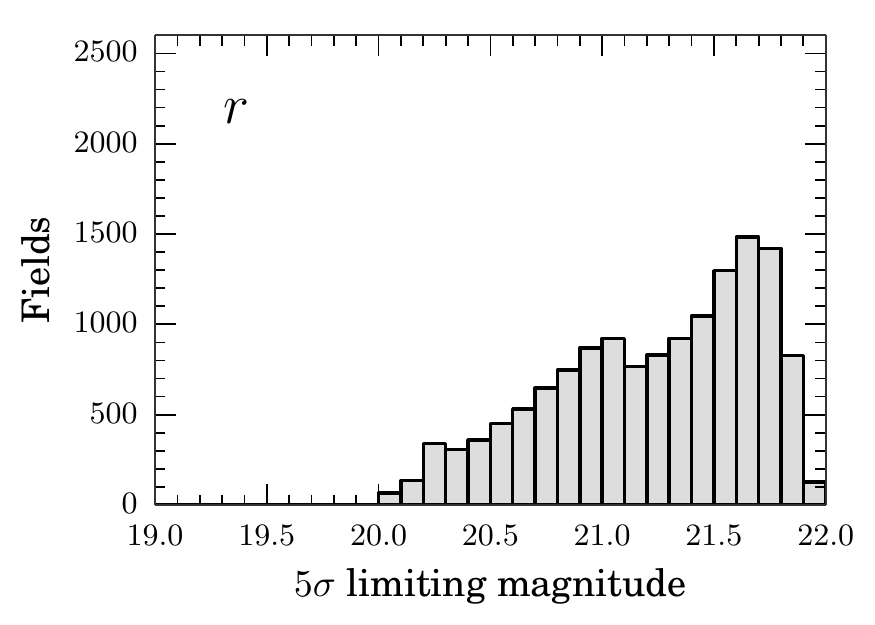} 
    \end{minipage}
    \begin{minipage}[b]{\linewidth}
        \includegraphics[width=\textwidth]{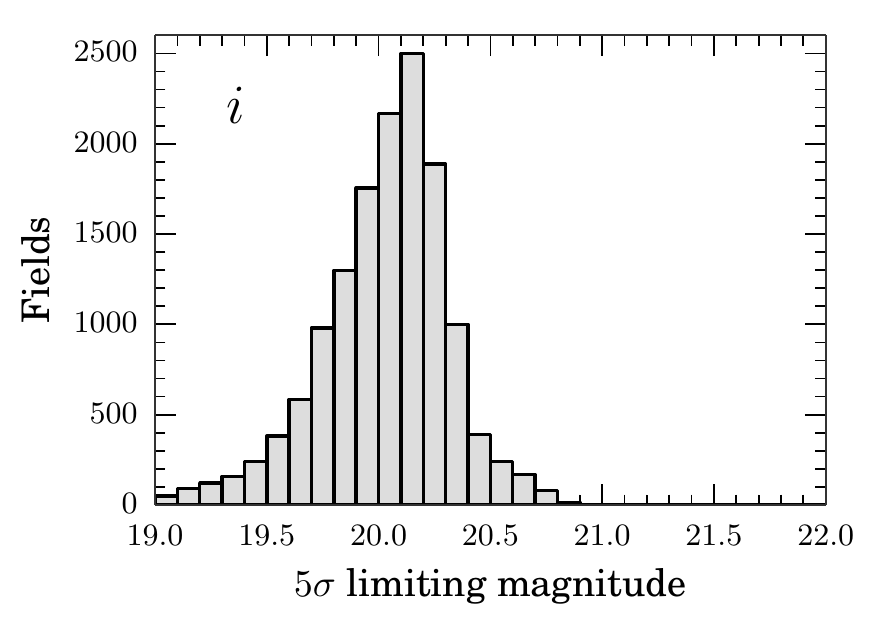} 
    \end{minipage}
    \begin{minipage}[b]{\linewidth}
        \includegraphics[width=\textwidth]{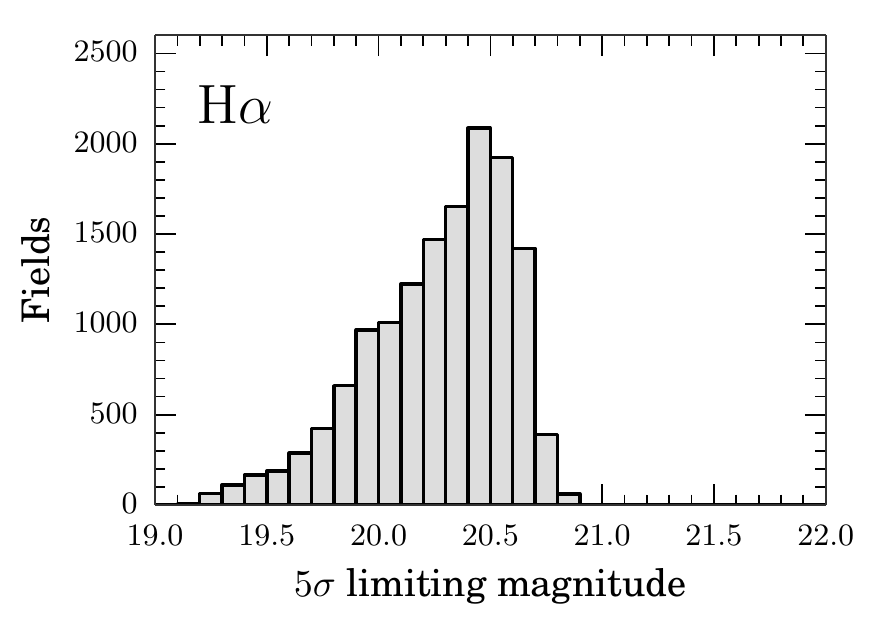} 
    \end{minipage}
    \caption{Distribution of the 5$\sigma$ limiting magnitude
             across all quality-approved survey fields
             for $r$ (top), $i$ (middle) and \ha\ (bottom).
             Fields with a limiting magnitude brighter than
             20th ($r$) or 19th (\ha, $i$) were rejected
             from the data release.
             The $r$-band depth is most sensitive 
             to the presence of the moon above the horizon: 
             this is the main reason for the wide, bi-modal character
             of its distribution.}
    \label{fig:depth}
\end{figure}

Observing time for IPHAS is obtained
on a semester-by-semester basis
through the open time allocation committees 
of the Isaac Newton Group of telescopes.
The survey is allocated specific observing dates
rather than particular observing conditions.
In consequence, data were acquired
under a large range of atmospheric conditions.
Data taken under unsuitable conditions
have been rejected using seven quality criteria,
which ensure a good and homogeneous level of quality
across the data release:

(1) \emph{Depth.} 
We discarded any exposures 
for which the $5\sigma$ limiting magnitude\footnote{
We defined the $5\sigma$ limiting magnitude
as the magnitude which a point source would have
if its flux equalled five times the level of the noise in the sky background.
The sky noise was estimated using a robust MAD estimator 
for noise scaled to equivalent Gaussian standard deviation,
ie. = MAD x 1.48,
after removing large scale sky background variations.
(MAD = Median of the Absolute Deviations about the median.)}
was brighter than 20th magnitude in the $r$-band
or brighter than 19th in $i$ or \ha.
Such data were typically obtained during poor weather or full moon.
Most observations were significantly better than these limits.
Fig.~\ref{fig:depth} presents the distribution of limiting magnitudes
for all quality-approved fields;
the mean depths and standard deviations are 
$21.2\pm0.5$ ($r$), $20.0\pm0.3$ ($i$) and $20.3\pm0.3$ (\ha).
The depth achieved depended
most strongly on the presence of the moon,
which was above the horizon during 62 per cent 
of the observations.
The great range in sky brightness this
produced is behind the wide and bi-modal shape
of the $r$-band limiting magnitude distribution 
(top panel in Fig.~\ref{fig:depth}).
In contrast, the depths attained in $i$ and \ha\ 
are less sensitive to moonlight, leading to
narrower magnitude limit distributions
(middle and bottom panels in Fig.~\ref{fig:depth}).
To a lesser extent, the wide spread in the $r$-band depth
is also explained by the shorter exposure time
that was used for this band during the first months of data-taking.

(2) \emph{Ellipticity.} 
The ellipticity of a point source,
defined as $e = 1 - b / a$ 
with $b$ the semi-minor and $a$ the semi-major axis,
is a morphological measure of the elongation
of the Point Spread Function (PSF).
It is expected to be zero (circular)
in a perfect noise-free imaging system,
but it is slightly non-zero in any real telescope data 
due to optical distortions, tracking errors and photon plus readout noise.
Indeed, it is worth noting
that IPHAS data have been collected
from unguided exposures that rely entirely
on the INT's tracking capability.
The mean ellipticity measured
in the data is $0.09\pm0.04$.
There have been sporadic episodes with higher ellipticities
due to mechanical glitches in the telescope tracking system.
As a result, 3 per cent of our images show an
average ellipticity across the detectors which is worse
than $e > 0.2$.
The inspection of these examples revealed no evidence
for degraded photometry up to ellipticities of 0.3.
We have excluded a small number of exposures
which exceeded $e = 0.3$.

\begin{figure}
    \begin{minipage}[b]{\linewidth}
        \includegraphics[width=\textwidth]{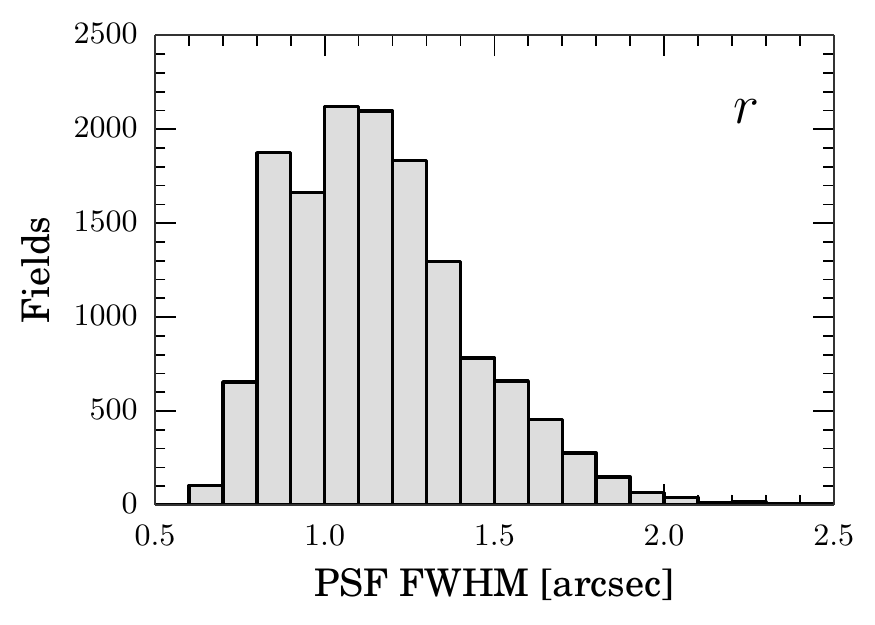} 
    \end{minipage}
    \begin{minipage}[b]{\linewidth}
        \includegraphics[width=\textwidth]{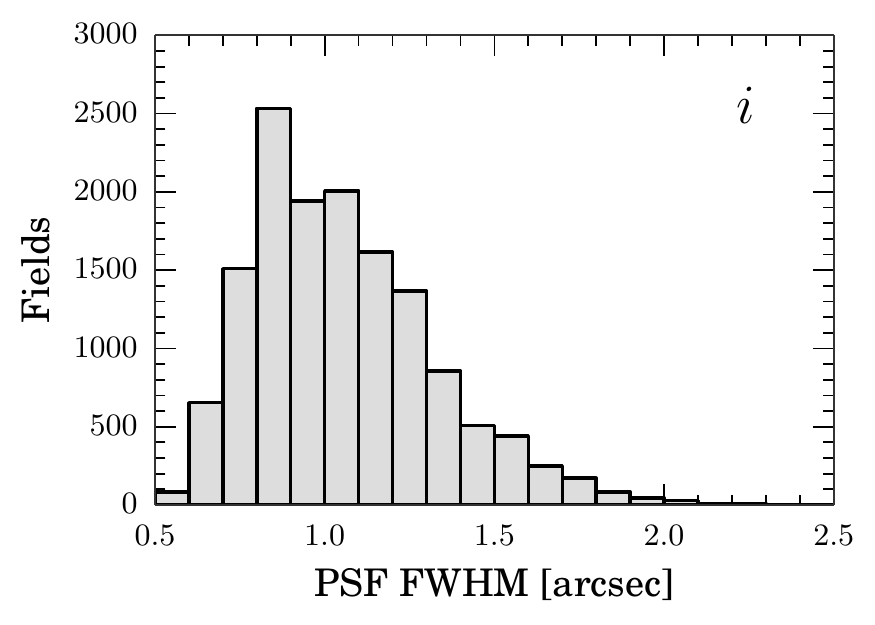} 
    \end{minipage}
    \begin{minipage}[b]{\linewidth}
        \includegraphics[width=\textwidth]{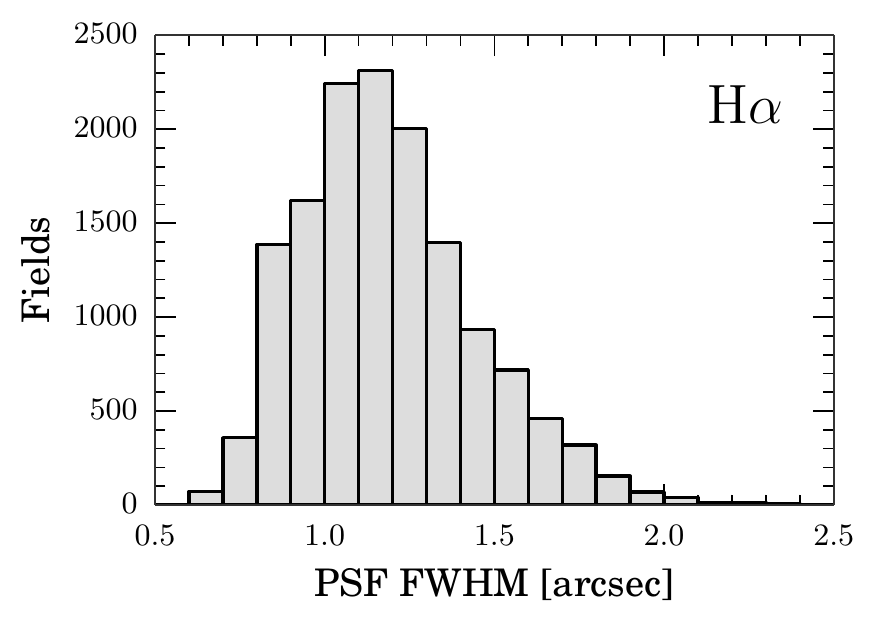} 
    \end{minipage}
    \caption{Distribution of the PSF FWHM
             for all the quality-approved fields
             included in the release,
             measured in $r$ (top), $i$ (middle) and \ha\ (bottom).
             The PSF FWHM measures the effective image resolution
             that arises from the combination of atmospheric and dome seeing.}
    \label{fig:seeing}
\end{figure}

(3) \emph{Seeing.} 
The original survey goal was to obtain data 
at a resolution better than 1.7~arcsec,
as evaluated by measuring the average PSF Full Width at Half Maximum (FWHM)
across the detectors.
This target is currently attained across 86 per cent of the footprint.
To increase the sky area offered by the data release slightly,
we have decided to accept data obtained with FWHM up to 2.5~arcsec.
Fig.~\ref{fig:seeing} presents the distribution
of the PSF FWHM for the approved fields.
In the $r$-band, 90 per cent is better than 1.5~arcsec,
50 per cent is better than 1.1~arcsec,
and 10 per cent is better than 0.8~arcsec.
In \S\ref{sec:catalogue} we will explain
that the photometry compiled in the source catalogue
is normally derived from the field with the
best-available seeing for a given object,
and that the FWHM measurement
is available as a column in the catalogue.

(4) \emph{Photometric repeatability.} 
The IPHAS field-pair observing strategy normally
ensures that every pointing is immediately followed 
by an offset pointing at a displacement
of $+$5~arcmin in Dec and $+$5~arcmin in RA.
This allows pairs of images to be checked 
for the presence of clouds or electronic noise.
To exploit this information,
the overlap regions of all field pairs were systematically cross-matched
to verify the consistency of the photometry
for stars observed in both pointings.
We automatically rejected field pairs
in which more than 2 per cent of the stars 
showed an inconsistent measurement at the level of 0.2~mag,
or more than 25 per cent were inconsistent at the level of 0.1~mag.
These limits were set empirically after inspecting
the images and photometry by eye.

(5) \emph{Visual examination.}
Images, colour mosaics,
and the associated photometric colour/magnitude diagrams
were inspected by a team of 20 survey members, 
such that each image in the data release 
was looked at by at least three different pairs of eyes.
Images affected by clouds or extreme levels of scattered moonlight
were flagged, investigated,
and excluded from the release 
if deemed necessary.

(6) \emph{Source density mapping.}
Spatial maps showing the number density of the detected sources
down to 20th magnitude were created to verify the health
of the data and to check for unexpected artefacts.
In particular, we created density maps
which showed the number of \emph{unique} sources
obtained by cross-matching the detection tables of
all three bands with a maximum matching distance of 1~arcsec.
This was effective for revealing
fields with an inaccurate astrometric solution in one of the bands,
which were subsequently corrected.

(7) \emph{Contemporaneous field data.} 
Finally, only exposures which are part of a sequence 
of three consecutive images of the same field
(\ha, $r$, $i$)
were considered for inclusion in the release. 
This ensures that the three bands for a given field
are observed contemporaneously --  
nearly always within 5 minutes of each other.
We note that the source catalogue details the exact epoch
at the start of each exposure
(columns \emph{rMJD}, \emph{iMJD}, \emph{haMJD}).

The above criteria were satisfied by at least one observing attempt
for 14,115 out of the 15,270 planned fields (92 per cent).
In some cases more than one successful attempt to observe
a field was available due to stricter
quality criteria being applied in the initial years of the survey.
In such cases, only the attempt 
with the best seeing and depth has been selected
for inclusion in the catalogue, in order  
to deliver the most accurate measurement at a single epoch.

We note that some of the excluded data may nevertheless be useful
for e.g. time-domain studies of bright stars.
The discarded data is
available through our website,
but will be ignored in the remainder of this work.

\section{Photometric calibration}
\label{sec:calibration}

Having obtained a quality-approved set of observations,
we now turn to the challenge of placing the data
onto a uniform photometric scale.

\subsection{Provisional nightly calibration}

For the purpose of providing an initial calibration 
of the $r$ and $i$ broadband fluxes,
photometric standard fields were observed every night.
The standards were chosen from a list based on 
the \cite{Landolt1992} and Stetson (http://cadcwww.dao.nrc.ca/standards) 
objects.
Two or three standard fields were observed 
during the evening and morning twilight,
and at intervals of 2-3 hours throughout the night.
The CASU pipeline automatically identified the observed standards 
and used them to determine a sigma-clipped average zeropoint \textsc{magzpt}
for each night and filter,
such that the number counts $DN$ 
in the pipeline-corrected CCD frames
relate to a magnitude $m$ as:
\begin{equation}
\begin{split}
   m  = & \textsc{magzpt} - 2.5 \log_{10}( DN / \textsc{exptime} ) \\
 &  - \textsc{extinct}\cdot(\textsc{airmass}-1) - \textsc{apcor} - \textsc{percorr},
\label{eqn:mag}
\end{split}
\end{equation}
where \textsc{exptime} is the exposure time in seconds,
\textsc{extinct} is the atmospheric extinction coefficient 
(set in the pipeline at 0.09 for $r$ and 0.05 for $i$ as representative
averages for the telescope site),
\textsc{airmass} is the normalised optical path length 
through the atmosphere and
\textsc{apcor} is a correction for the flux
lost outside of the aperture used
(we adopt a 2\farcs 3-diameter circular aperture by default).
Finally, \textsc{percorr} is a term used to correct 
for the small difference in internal gain
computed using the relatively blue twilight flats
compared to the much redder typical astronomical objects. 
It is estimated by making a robust average of the dark sky levels 
measured on each detector during an observing run
(the correction is $0.01\pm0.01$ on average in $i$ and
$0.00\pm0.00$ in $r$ and \ha).
All these quantities correspond to header keywords
in the FITS files produced by the CASU pipeline.

The broadband zeropoints were determined such that the resulting magnitude system
refers to the spectral energy distribution (SED) of Vega 
as the zero colour object. 
Colour equations were used to transform between the IPHAS passbands 
and the Johnson-Cousins system 
of the published standard star photometry.
The entire procedure has been found to deliver zeropoints which 
are accurate at the level of 1--2 per cent 
in stable photometric conditions \citep{Gonzalez-Solares2011}.

Unlike the broadbands, 
standard-star photometry is not available in the literature 
for the \ha\ passband
and hence there is no formally recognised flux scale for it.
We can specify here, however, 
that the detected flux for Vega in the
IPHAS \ha\ filter is 3.14 magnitudes
less than the flux captured by the much broader
$r$ band \citep{Gonzalez-Solares2008}.
Hence to assure that $(r-\ha)=0$ for Vega,
we set the zeropoint for the narrowband to be:
\begin{equation}
\textsc{magzpt}_{H\alpha} = \textsc{magzpt}_r - 3.14.
\label{eqn:zpha}
\end{equation}
For reference, Table~\ref{tbl:flux} details the flux of Vega
in the IPHAS filter system.
Data on the throughput curves of the bands can be obtained from
the Isaac Newton Group website\footnote{http://catserver.ing.iac.es/filter/list.php?instrument=WFC where the
filters are named WFCH6568, WFCSloanR and WFCSloanI}.

\begin{table}
    \caption{Mean monochromatic flux of Vega in the IPHAS filter system,
             defined as 
             $\langle f_\lambda\rangle~=~\int{f_\lambda(\lambda) S(\lambda) \lambda d \lambda} / \int{S(\lambda)\lambda d \lambda}$,
             where $S(\lambda)$ is the photon response function 
             (which includes atmospheric transmission,
             filter transmission
             and CCD response)
             and $f_\lambda(\lambda)$ is the CALSPEC SED for Vega
             \citep{Bohlin2014}.
             For reference, we also provide the 
             filter equivalent width EW~=~$\int S(\lambda) d\lambda$,
             the mean photon wavelength 
             $\lambda_0~=~\int{S(\lambda)\lambda d\lambda} / \int{S(\lambda)d\lambda}$,
             and the pivot wavelength
             $\lambda_p~=~\sqrt{\int{S(\lambda)\lambda d\lambda} / \int{\frac{S(\lambda)}{\lambda} d\lambda}}$.
             These notations follow the definitions by \citet{Bessell2012}.
             After multiplying $\langle f_\lambda\rangle$ by the EW,
             we find that the detected flux for Vega in \ha\ is 3.14~magnitudes less
             than that received in $r$.
          }
    \label{tbl:flux}
    \begin{center}
    \begin{tabular}{lccrc}
    \toprule
     Filter  & $\langle f_\lambda \rangle$  & \multicolumn{1}{c}{EW} &$\lambda_0$ & $\lambda_p$    \\
             & [erg\,cm$^{-2}$\,s$^{-1}$\,\AA$^{-1}$] & [\AA] & [\AA] & [\AA] \\
    \midrule
$r$ &  $2.47 \times 10^{-9}$  &  785.6 &  6223  &   6211   \\
\ha &  $1.81 \times 10^{-9}$ &   59.6 &  6568  &   6568    \\
$i$  &  $1.30 \times 10^{-9}$  &  759.9  &  7674  &   7661   \\
    \bottomrule
    \end{tabular}
    \end{center}
\end{table}

\subsection{Global re-calibration}

\begin{figure}
\captionsetup[subfigure]{labelformat=empty}
\begin{subfigure}[b]{\linewidth}
\centering
\includegraphics[width=\textwidth]{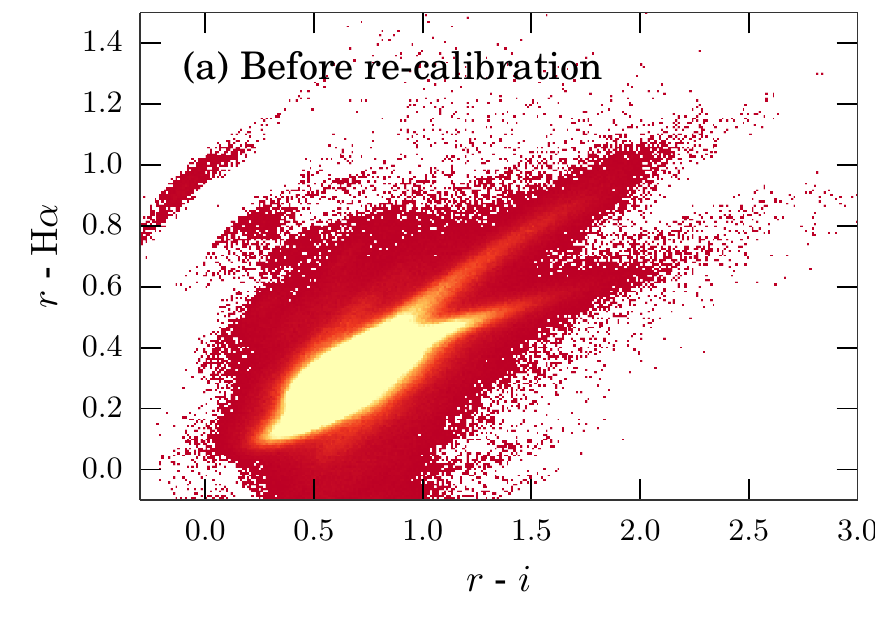}
\caption{}
\label{fig:ccd_before}
\end{subfigure}
\begin{subfigure}[b]{\linewidth}
\centering
\includegraphics[width=\textwidth]{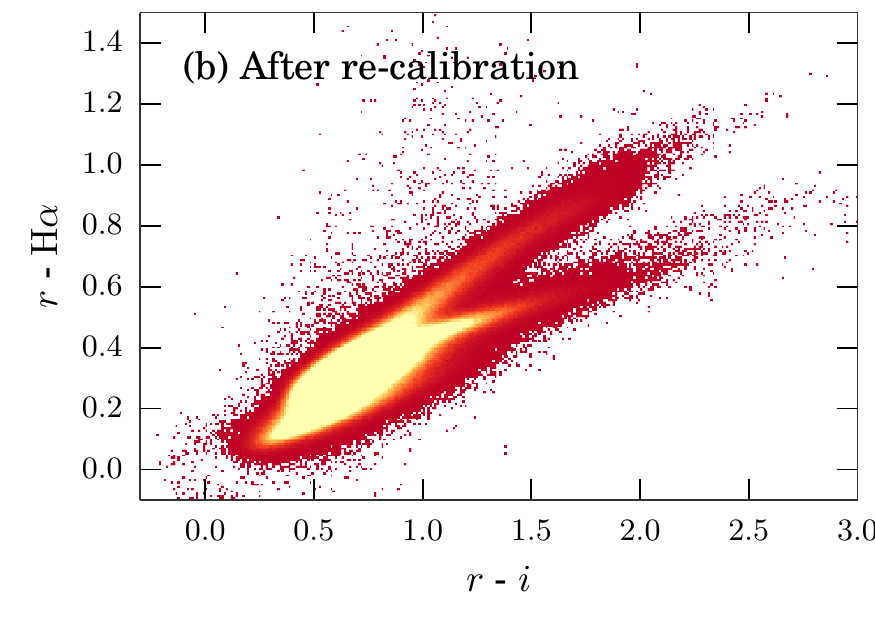}
\caption{}
\label{fig:ccd_after}
\end{subfigure}
\caption{IPHAS $(r-\ha,\ r-i)$ colour-colour diagram
         covering an area of 400~deg$^2$,
         shown before (panel a) and after (panel b)
         re-calibration.
         Both figures were created by combining the stars
         detected across all 2,801 quality-approved fields
         which are located towards the Galactic anti-centre
         ($160\degr<\ell<200\degr$).
         The diagrams are plotted as 2D-histograms
         which show the density of sources
         in bins of 0.01-by-0.01~mag;
         bins containing 1 to 200 sources are coloured red,
         while bins containing more than 400 sources are bright yellow.
         The diagrams include all stars
         brighter than $r<18$ which were
         classified by the pipeline as `a10point'
         (indicating a high-significance point source detection with
         accurate photometry in all bands,
         to be explained in \S\ref{sec:qualitycriteria}).
         The objects which are seen to fall above the locus of stars
         \emph{after} re-calibration 
         are likely to be genuine \ha\ emission-line objects.}
\end{figure}

Despite the best efforts made to obtain a nightly calibration,
large surveys naturally possess field-to-field variations
due to atmospheric changes during the night
and imperfections in the pipeline or the instrument
(e.g. the WFC is known to suffer from sporadic errors
in the timing of exposures).
This is demonstrated in Fig.~\ref{fig:ccd_before},
where we show the combined colour-colour diagram
for nearly 3,000 fields across an area of 400~deg$^2$.
The main locus of stars is poorly defined
in the diagram due to the presence of
incorrectly calibrated fields,
which need to be corrected
during a global re-calibration step.
The application of such a procedure,
to be explained below,
has revealed that the error in our initial nightly calibration
exceeded 0.1~mag in 12 per cent of the fields,
and even exceeded 0.5~mag in 0.7 per cent.
Fig.~\ref{fig:ccd_after}
demonstrates the improvement 
in the colour-colour diagram
after re-calibrating.

Notable past examples of surveys which required
global re-calibration include 
2MASS \citep{Nikolaev2000},
SDSS \citep{Padmanabhan2008}
and the Panoramic Survey Telescope 
And Rapid Response System survey \citep[Pan-STARRS;][]{Schlafly2012},
which all achieved photometry 
that is globally consistent to within 0.01--0.02~mag
after re-calibration.

Surveys which observe identical stars at different epochs
can use the repeat measurements to ensure a homogeneous calibration.
For example, 2MASS attained its global calibration
by observing two standard fields each hour,
allowing zeropoint variations to be tracked 
over short timescales \citep{Nikolaev2000}.
Alternatively, the SDSS and PanSTARRS surveys could benefit
from revisiting regions in their footprint to 
carry out a so-called \emph{ubercalibration}\footnote{`ubercalibration'
refers to the name of the code used to re-calibrate SDSS photometry. 
It is an anglicised version of the German word `\"uberkalibration',
which was reportedly chosen because the initial authors, Schlegel and Finkbeiner, both
have German-sounding names \citep{Finkbeiner2010}.} procedure,
in which repeat measurements of stars in different nights
are used to fit the calibration parameters
\citep{Ivezic2007,Padmanabhan2008,Schlafly2012}.

Unfortunately these schemes cannot be applied
directly to IPHAS
for two reasons. 
First, the survey was carried out 
in competitively-allocated observing time on a
common-user telescope, 
rendering the 2MASS approach 
of observing standards at a high frequency
prohibitively expensive (it does not help that 
standard fields are very scarce within the Galactic Plane).
Second, IPHAS is not specified as a variability survey,
with the result that stars are not normally observed
at more than one epoch,
unless they happen to fall within a narrow overlap region 
between two neighbouring field pairs.

We have found the information contained
in our narrow overlap regions to be insufficient
to constrain the calibration parameters well enough.
This is because photometry at the extreme edges of the WFC
-- where neighbouring field pairs overlap -- 
is prone to systematics at the level of 1--2~per cent.
The cause of these errors is thought to include 
the use of twilight sky flats in the pipeline,
which are known to be imperfect for calibrating stellar photometry 
due to stray light and vignetting \citep[e.g.][]{Manfroid1995}.
Moreover, the illumination correction in the overlap regions
is more affected by a radial geometric distortion in the WFC,
which causes the pixel scale to increase as the edges 
are approached \citep{Gonzalez-Solares2011}.
Although these systematics are reasonably small within a single field,
they can accumulate during a re-calibration process,
causing artificial zeropoint gradients across the survey
unless controlled by other external constraints.

For these reasons, we have not depended
on an ubercalibration-type scheme alone,
but have opted to involve an external reference survey
-- where available --
to bring the majority of our data onto a homogeneous calibration.

\subsubsection{Correcting zeropoints using APASS}

\begin{figure*}
    \includegraphics[width=\textwidth]{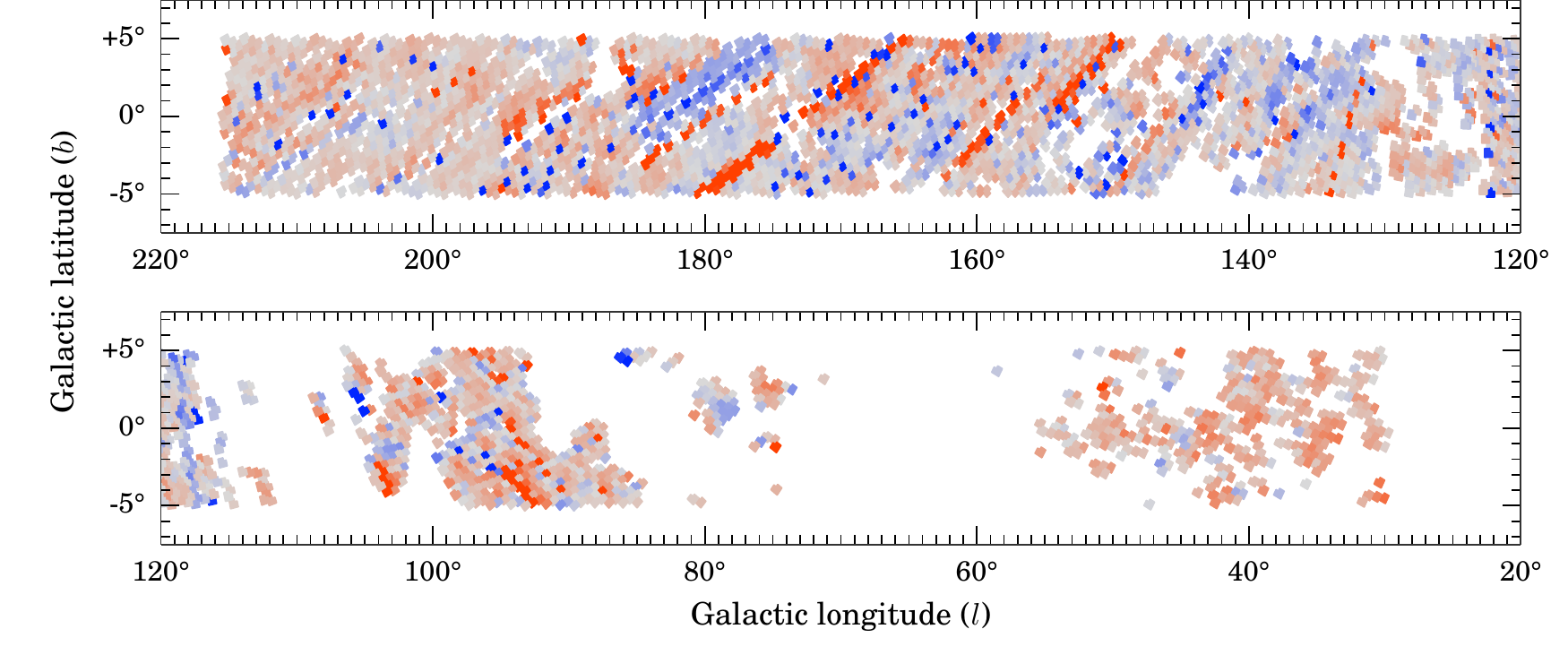} 
    \includegraphics[width=\textwidth]{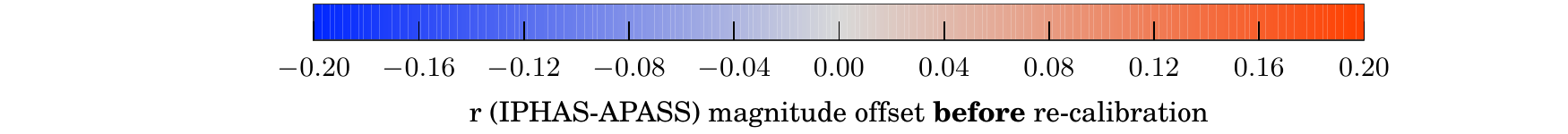} 
    \caption{Median magnitude offset in the $r$ band between IPHAS and APASS,
             plotted on a field-by-field basis
             prior to the re-calibration procedure.
             Each square represents the footprint of an IPHAS field
             which contains at least 30 stars with a counterpart
             in the APASS DR7 catalogue.
             The colours denote the median
             IPHAS-APASS magnitude offset in each field,
             which was computed after applying the APASS-to-IPHAS
             transformation to the APASS magnitudes (Eqn.~\ref{eqn:apass_r}).
             For clarity, we do not show the fields at the offset positions.}
        \label{fig:apass_r}
    \vspace{1cm}
    \includegraphics[width=\textwidth]{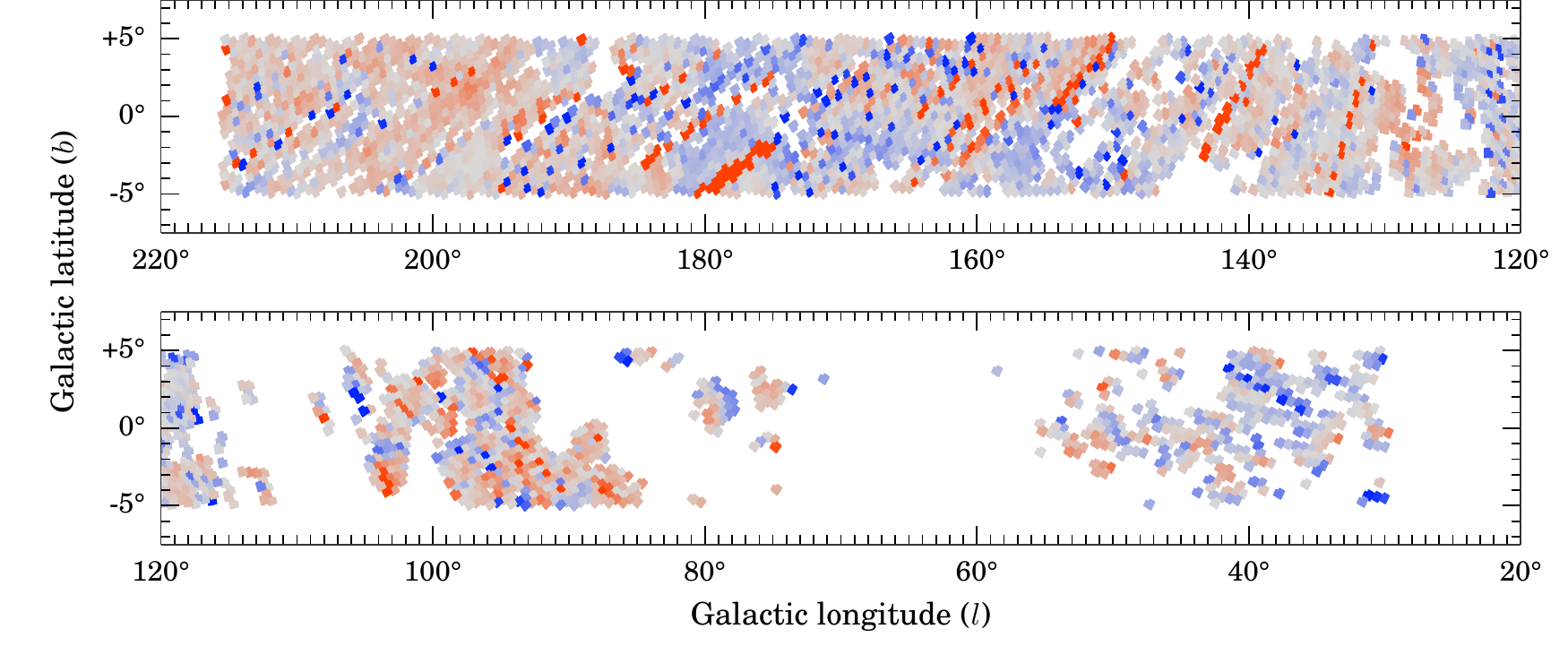} 
    \includegraphics[width=\textwidth]{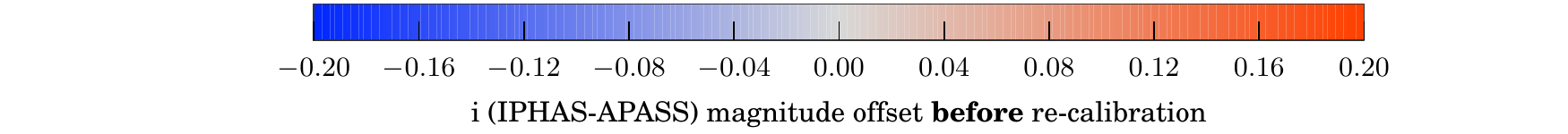} 
    \caption{Same as Fig.~\ref{fig:apass_r} for the $i$-band.}
    \label{fig:apass_i}
\end{figure*}

We have been able to benefit from APASS
(http://www.aavso.org/apass)
to bring most of the survey 
onto a uniform scale.
Since 2009,
APASS has been using two 20~cm-astrographs
to survey the entire sky down to $\sim$17th magnitude
in five filters which include Sloan $r$ and $i$ \citep{Henden2012}.
The most recent catalogue available 
at the time of preparing this work was APASS DR7,
which provides a good coverage across $\sim$half
of the IPHAS footprint.
The overlap regions are shown in Fig.~\ref{fig:apass_r}.
The photometric accuracy of APASS is currently estimated 
to be at the level of 3 per cent,
which is significantly better 
than the original nightly calibrations of IPHAS
which are only accurate to $\sim$10 per cent
when compared to APASS (Table~\ref{tbl:offsets_before}).
APASS achieves its uniform accuracy 
by measuring each star at least two times in photometric conditions,
along with ample standard fields,
benefiting from the large 3-by-3 degrees field of view of its detectors.

With the aim of bringing IPHAS to a similar accuracy of 3 per cent,
we used the APASS catalogue to identify and adjust the calibration of all IPHAS fields 
which showed a magnitude offset larger than 0.03~mag against APASS.  Experience of re-running
the calibration and testing the results showed us that it was inadvisable to tune more
finely the match for IPHAS data obtained in what were generally the most photometric
nights.  To this end,
the $r$- and $i$-band detection tables of each IPHAS field
were cross-matched against the APASS DR7 catalogue 
using a maximum matching distance of 1~arcsec.
The magnitude range was limited to
$13<r_{\rm APASS}<16.5$ and $12.5<i_{\rm APASS}<16.0$
in order to avoid sources 
brighter than the IPHAS saturation limit on one hand, 
and to avoid sources near the faint detection limit of APASS 
on the other.

The resulting set of 220,000 cross-matched stars were then used 
to derive APASS-to-IPHAS magnitude transformations
using a linear least-squares fitting routine, 
which iteratively removed $3\sigma$-outliers to improve the fit.
The solution converged to:
\begin{align} 
r_{\rm IPHAS} = r_{\rm APASS} - 0.121 + 0.032(r-i)_{\rm APASS} \label{eqn:apass_r} \\
i_{\rm IPHAS} = i_{\rm APASS} - 0.364 + 0.006(r-i)_{\rm APASS} \label{eqn:apass_i}
\end{align}
The root mean square (rms) residuals of these transformations 
are 0.041 and 0.051, respectively.
The small colour terms in the equations
indicate that the IPHAS and APASS broadband filters 
are very similar.
The transformations include a large fixed offset,
but this is simply due to the fact that 
APASS magnitudes are given in the AB system
and IPHAS uses Vega-based magnitudes.
Separate transformations were derived for sightlines 
with varying extinction properties to investigate the robustness
of the transformations with respect to different reddening regimes.
This sensitivity was found to be insignificant, in keeping with
the scarcity of heavily-reddened objects at $r<16$.

Having transformed APASS magnitudes into the IPHAS system,
we then computed the median magnitude offset 
for each field which contained at least 30 cross-matched stars.
This was achieved for 48 per cent of our fields (shown in Figs.~\ref{fig:apass_r} and \ref{fig:apass_i}).
The offsets follow a near-Gaussian distribution
with mean and sigma $0.014\pm0.104$~mag in $r$
and $0.007\pm0.108$~mag in $i$ (Table~\ref{tbl:offsets_before}).
A total of 4,596 IPHAS fields showed a median offset
exceeding $\pm$0.03~mag in either $r$ or $i$
when compared to APASS.

\begin{table}
    \caption{Magnitude offsets for objects
             cross-matched between IPHAS and APASS/SDSS
             \emph{before} the global re-calibration was carried out.
             We characterise the distribution of the offsets,
             which is approximately Gaussian in each case,
             by listing the mean and the standard deviation values.
			 We remind the reader that transformations were applied
             to the APASS and SDSS magnitudes to bring them into the
             Vega-based IPHAS system prior to computing the offsets.
             }
    \label{tbl:offsets_before}
    \begin{center}
        \begin{tabular}{lcc}
            \toprule
            {\bf Before re-calibration} & Mean & $\sigma$  \\
            \midrule
            $r$ (IPHAS - APASS) & +0.014 & 0.104 \\
            $i$ (IPHAS - APASS) & +0.007 & 0.108 \\
            $r$ (IPHAS - SDSS) & +0.016 & 0.088 \\
            $i$ (IPHAS - SDSS) & +0.010 & 0.089 \\
            \bottomrule
       \end{tabular}
       \vspace{1cm}
       \caption{Same as Table~\ref{tbl:offsets_before},
                but computed \emph{after} the global re-calibration
                was carried out. The mean and standard deviation values
                of the offsets have improved significantly.}
        \label{tbl:offsets_after}
            \begin{tabular}{lcc}
                \toprule
                {\bf After re-calibration} & Mean & $\sigma$ \\
                \midrule
                $r$ (IPHAS - APASS) & +0.000 & 0.011\\
                $i$ (IPHAS - APASS) & +0.000 & 0.011 \\
                $r$ (IPHAS - SDSS)  & -0.001 & 0.029\\
                $i$ (IPHAS - SDSS) & -0.002 & 0.032 \\
                \bottomrule
            \end{tabular}
        \end{center}
\end{table}

We then applied the most important step
in our re-calibration scheme,
which is to adjust the provisional zeropoints
of these 4,596 aberrant fields
such that their offset is brought to zero.
This allowed the mean IPHAS-to-APASS offset 
to be brought down to $0.000\pm0.011$~mag in both $r$ and $i$
(Table~\ref{tbl:offsets_after}).
The procedure of fitting magnitude transformations and
correcting the IPHAS zeropoints was repeated a few times to ensure 
convergence, which was closely approached after the first iteration.

\subsubsection{Adjusting fields not covered by APASS}

At the time of writing, the APASS catalogue did not provide 
sufficient coverage for 7,359 of our fields.
Fortunately, these fields are located mainly 
at low Galactic longitudes (cf. Figs.~\ref{fig:apass_r} and \ref{fig:apass_i}),
which were typically observed during the summer months
when photometric conditions are more prevalent at the telescope.
These remaining fields have nevertheless 
been brought onto the same uniform scale 
by employing an ubercalibration-style scheme,
which minimises the magnitude offsets between stars
located in the overlap regions between neighbouring fields.

An algorithm for achieving this minimisation
has previously been described by~\citet{Glazebrook1994}.
In brief, there are two fundamental quantities to be
minimised between each pair of overlapping exposures, 
denoted by the indices $i$ and $j$. 
First, the mean magnitude difference between stars in the overlap
region $\Delta_{ij}=\langle m_i-m_j\rangle=-\Delta_{ji}$ is a local
constraint. 
Second, to ensure the solution does not stray far 
from the existing calibration, 
the difference in zeropoints 
$\Delta\mathrm{ZP}_{ij}=-\Delta\mathrm{ZP}_{ji}$ 
between each pair of exposures must also be minimised.

Minimisation of these two quantities is a linear least squares problem 
because the magnitude $m$ depends linearly on the ZP (Eqn.~\ref{eqn:mag}).
Hence we can find the ZP shift to be applied to each field 
by minimising the sum:
\begin{equation}
   S = \sum_{i=1}^N \sum_{j=1}^N w_{ij} \theta_{ij} (\Delta_{ij} + a_i - a_j)^2,
   \label{eqn:chi2}
\end{equation}
where $i$ denotes an exposure, 
$j$ an overlapping exposure, 
$N$ the number of exposures,
$a_i$ the ZP to solve for,
and $a_j$ the ZP of an overlapping field ($\Delta\mathrm{ZP}_{ij}=a_i-a_j$). 
$w_{ij}$ are weights set to the inverse square of the uncertainty in $\Delta_{ij}$,
and $\theta_{ij}$ is an overlap function 
equal to either 1 if exposures $i$ and $j$ overlap or 0 otherwise. 
Solving for $a_i$ is equivalent to solving $\partial
S/\partial a_i=0$, which gives the matrix equation:
\begin{equation}
   \sum_{j=1}^N A_{ij} a_j = b_j,
   \label{eqn:matrix}
\end{equation}
where 
\begin{eqnarray}
   A_{ij} &=& \delta_{ij} \sum_{k=1}^N w_{jk}\theta_{jk} - w_{ij} \theta_{ij},\label{eqn:aij}\\
   b_i &=& \sum_{j=1}^N w_{ij} \theta_{ij}\Delta_{ji} = -\sum_{j=1}^N w_{ij} \theta_{ij}\Delta_{ij}.\label{eqn:bi}
\end{eqnarray}

We enforce a strong external constraint
on the solution by keeping the zeropoint fixed 
for the fields which have already been compared
and calibrated against APASS.
We hereafter refer to these fields as \emph{anchors}.
It is asserted that the zeropoints $a_i$ of the anchor fields 
are known and not solved for.  However they do appear in the vector 
$b_j$ as constraints.  In addition to the APASS-based anchors, 
we selected 3,273 additional anchor fields by hand
to provide additional constraints in regions not covered by APASS.
These extra anchors were deemed to have accurate zeropoints 
based on 
(i) the information contained in the observing logs,
(ii) the stability of the standard star zeropoints during the night, and
(iii) photometricity statistics provided by the Carlsberg Meridian Telescope,
which is located $\sim$500~m from the INT.

We then solved Eqn.~\ref{eqn:matrix} for the $r$ and $i$ bands
separately using the least-squares routine 
in Python's {\sc scipy.sparse} module for sparse matrix algebra.
This provided us with corrected zeropoints for the remaining fields,
which were shifted on average by $+0.02\pm0.11$ in $r$ 
and $+0.01\pm0.12$ in $i$ compared to their provisional calibration.

We then turned to the global calibration of the \ha\ data.
It is not possible to re-calibrate the narrowband 
in the same way as the broadbands,
because the APASS survey does not offer \ha\ photometry.
We can reasonably assume, however,
that the corrections required for $r$ and \ha\ are identical,
much of the time, because the IPHAS data-taking pattern ensured 
that a field's \ha\ and $r$-band exposures
were taken consecutively,
albeit separated by a $\sim$30~s read-out time.
Hence, we have corrected the \ha\ zeropoints 
by re-using the zeropoint adjustments that were derived for the $r$ band
in the earlier steps.
An exception was made for 3,101 fields
for which our quality-control routines revealed
strong zeropoint variations during the night,
suggesting non-photometric conditions.
In these cases, the \ha\ zeropoints
were adjusted by solving Eqn.~\ref{eqn:matrix}
rather than by simply applying Eqn.~\ref{eqn:zpha}.

\subsection{Testing the calibration against SDSS}

Having re-calibrated all fields to the expected APASS accuracy of 3 per cent,
we turned to a different survey, SDSS Data Release 9 \citep{Ahn2012},
to validate the results.
SDSS DR9 includes photometry across several strips at low
Galactic latitudes,
which were observed as part of 
the Sloan Extension for Galactic Understanding and Exploration \citep[SEGUE;][]{segue}.
These strips provide data across 18 per cent of the fields in our data release.
We cross-matched the IPHAS fields against the subset of
objects marked as reliable stars in the SDSS catalogue\footnote{
We used the CasJobs facility located at http://skyserver.sdss3.org/CasJobs
to obtain photometry from the SDSS {\sc photoprimary} table 
with criteria {\sc type = star}, {\sc clean = 1} and {\sc score $>$ 0.7}.}
in much the same way as for APASS,
with the difference of selecting from the  fainter magnitude ranges of 
$15<r_{\rm SDSS}<18.0$ and $14.5<i_{\rm SDSS}<17.5$.
This provided us with a set of 1.2~million cross-matched stars.

Colour transformations were again obtained using a sigma-clipped linear least squares fit:
\begin{eqnarray}
r_{\rm IPHAS} = r_{\rm SDSS} - 0.093 - 0.044(r-i)_{\rm SDSS} \label{eqn:sdss_r} \\
i_{\rm IPHAS} = i_{\rm SDSS} - 0.318 - 0.095(r-i)_{\rm SDSS}. \label{eqn:sdss_i}
\end{eqnarray}
The rms residuals of these transformations are 0.045 and 0.073, respectively.
The equations are similar to the ones
previously determined for APASS,
although the colour terms are slightly larger.
The throughput curve of the SDSS $i$-band filter 
appears to be somewhat more sensitive at longer wavelengths
than both the IPHAS and APASS filters.

These global transformations were deemed adequate
for the purpose of validating our calibration in a statistical sense.
Separate equations were derived towards different sightlines
to investigate the effects of varying reddening regimes.
The colour term was found 
to show some variation towards weakly reddened areas,
where different stellar populations are observed.
The vast majority of red objects in the global sample
are those in highly reddened areas, however,
which agree well with the global transformations
and dominate the statistical appraisal of our calibration.

\begin{figure*}
    \includegraphics[width=\textwidth]{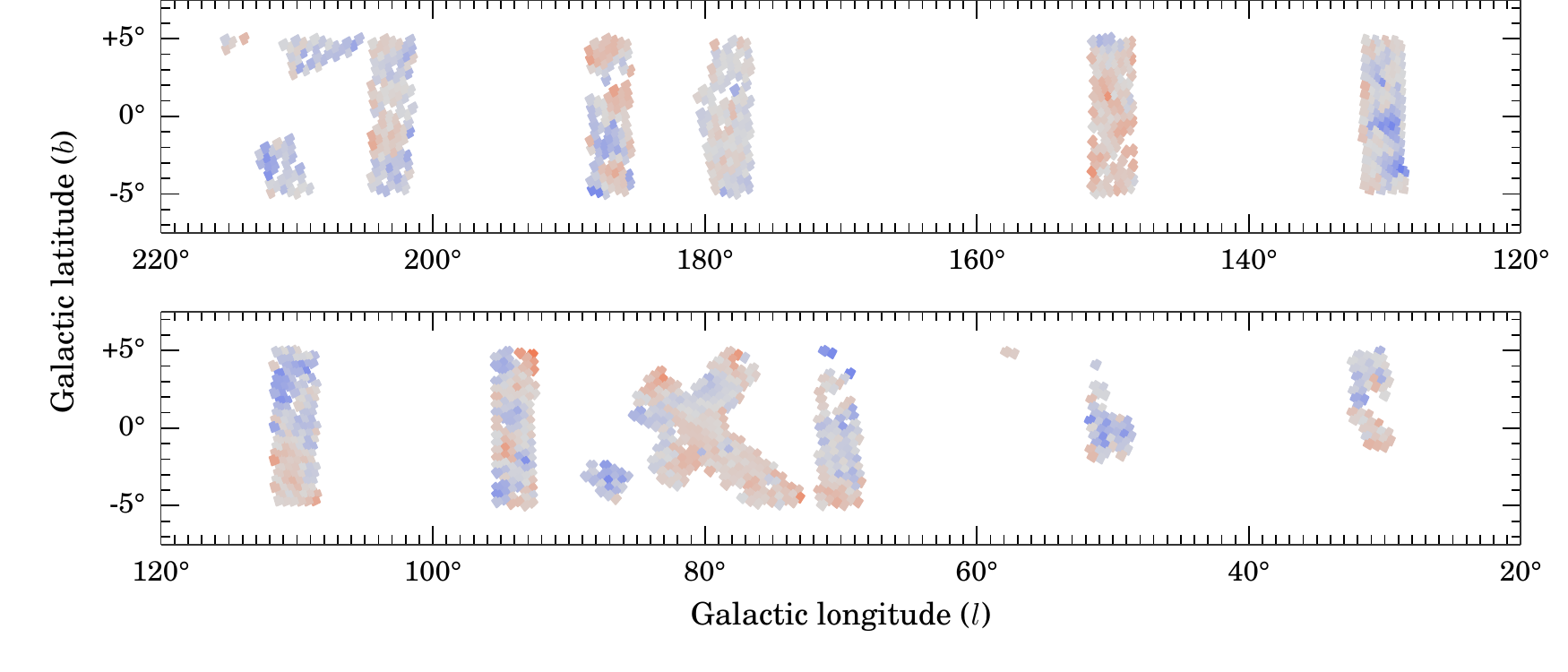}
    \includegraphics[width=\textwidth]{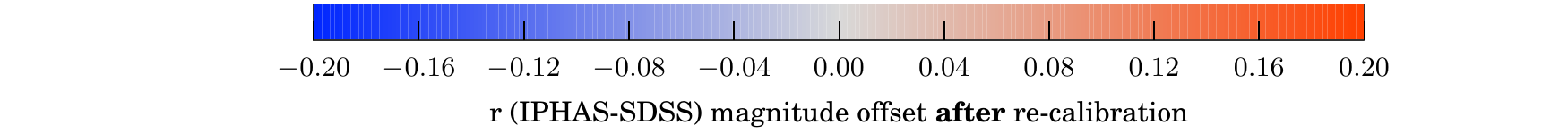} 
    \caption{Median magnitude offset between IPHAS and SDSS in the $r$ band
             \emph{after} the re-calibration procedure using APASS was applied.
             Each square represents the footprint of an IPHAS field
             which contains at least 30 stars
             with a counterpart in the SDSS DR9 catalogue.
             The colours denote the median IPHAS-SDSS magnitude offset
             in each field,
             which was computed after applying the SDSS-to-IPHAS
             transformation to the SDSS magnitudes (Eqn.~\ref{eqn:sdss_r}).}
    \label{fig:sdss_r}
    \vspace{1cm}
    \includegraphics[width=\textwidth]{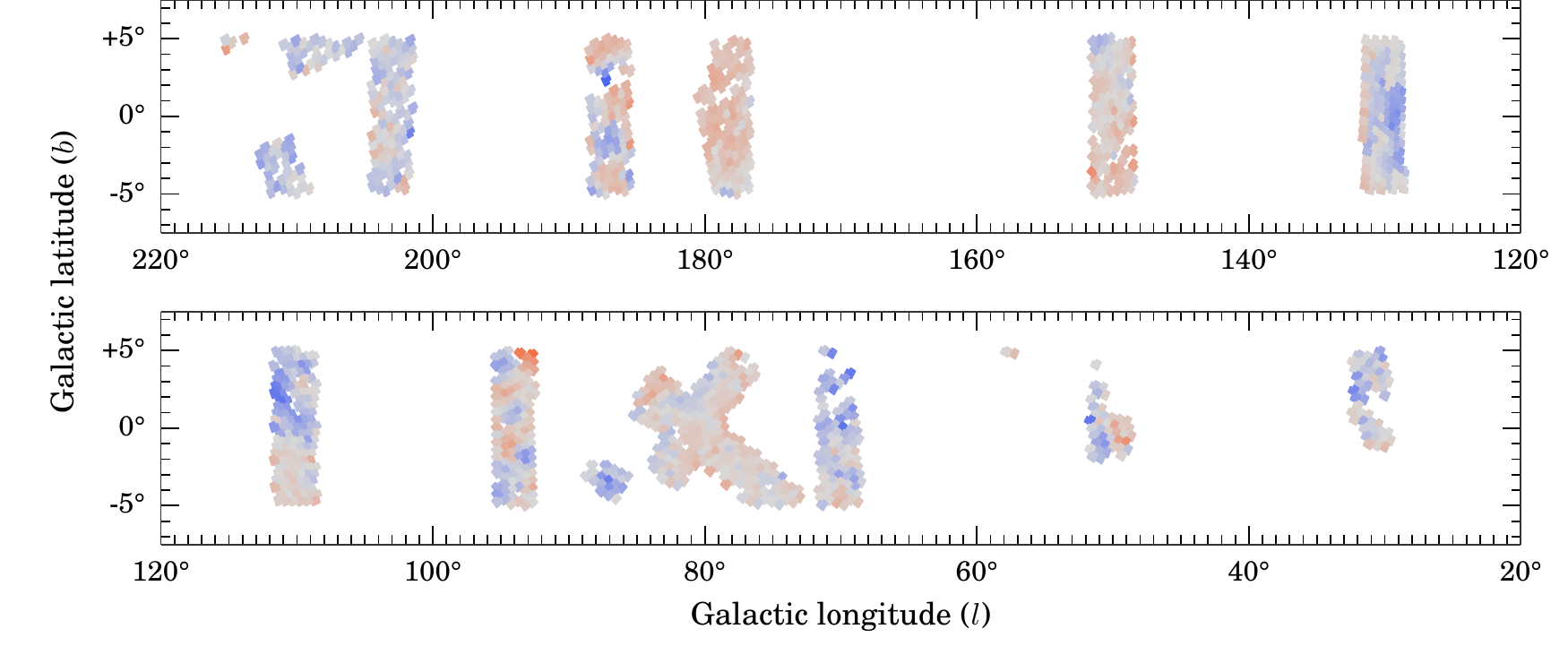}
    \includegraphics[width=\textwidth]{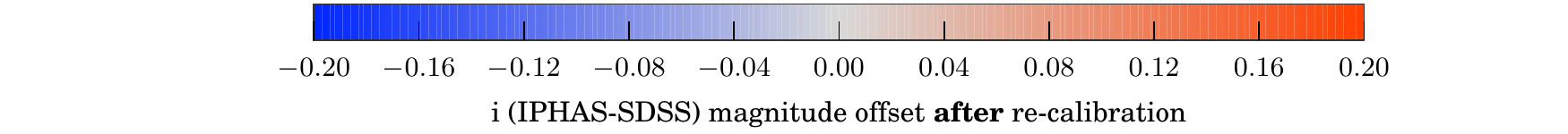} 
    \caption{Same as Fig.~\ref{fig:sdss_r} for the $i$-band.}
    \label{fig:sdss_i}
\end{figure*}

Having transformed SDSS magnitudes into the IPHAS system,
we then computed the median magnitude offset for each IPHAS field
which contained at least 30 objects with a cross-matched counterpart
in SDSS.  This was the case for 2,602 fields.
The median offsets for each of these fields
are shown in Figs.~\ref{fig:sdss_r} and \ref{fig:sdss_i}.
Importantly, the mean offset and standard deviation found 
is $-0.001\pm0.029$~mag in $r$
and $-0.002\pm0.032$~mag in $i$ (Table~\ref{tbl:offsets_after}).
In comparison, offsets computed in the identical way
\emph{before} carrying out the re-calibration showed means
of $+0.016\pm0.088$~mag in $r$ and $+0.010\pm0.089$~mag in $i$ (Table~\ref{tbl:offsets_after}).
We conclude that our re-calibration procedure has
been successful in improving the
uniformity of the calibration
by a factor three (i.e. from $\sigma=0.088$ to $\sigma=0.029$ in $r$),
and as such has achieved our aim of bringing the
accuracy to the aimed level of 0.03~mag.

The SDSS comparison revealed a number of fields where the offsets
exceeded 0.05~mag (523 fields) or even 0.1~mag (18 fields).
This pattern of outliers is consistent with the tails of a Gaussian distribution
with $\sigma=0.03$.  Furthermore, it should not
be forgotten that both the SDSS and APASS calibrations are approximations to
perfection and will not be entirely free of anomalies.  Indeed as we worked, 
we noticed the occasional unsurprising examples of inconsistency between these 
two surveys.

\section{Source catalogue generation}
\label{sec:catalogue}

Having obtained a quality-checked 
and re-calibrated data set, 
we now turn to the challenge
of transforming the observations 
into a user-friendly catalogue.
The aim is to present
the best-available information for each unique source
in a convenient format,
including flags to warn about quality issues 
such as source blending and saturation.
Compiling the catalogue involved four steps:
\begin{enumerate}
\item the single-band detection tables 
produced by the CASU pipeline 
were augmented with new columns
and warning flags;
\item the detection tables were merged into multi-band field catalogues;
\item the overlap regions of the field catalogues 
were cross-matched to flag duplicate (repeat) detections 
and identify the primary (best) detection 
of each unique source; and
\item these primary detections
were compiled into the final source catalogue.
\end{enumerate}
Each of these four steps are explained next.

\subsection{User-friendly columns and warning flags}

Enhancement of the detection tables by creating new columns
is the necessary first step because the tables generated by the CASU pipeline
refer to source positions in pixel coordinates, to photometric measures 
in number counts, and so on, rather than in common astronomical units.
To transform these data into user-friendly quantities,
we have largely adopted the units and naming conventions
which are in use at the
Wide Field Camera (WFCAM) Science Archive \citep[WSA;][]{Hambly2008}
and the 
Visible and Infrared Survey Telescope for Astronomy (VISTA)
Science Archive \citep[VSA;][]{Cross2012}.
These archives curate the high-resolution near-infrared photometry from both
the United Kingdom Infrared Telescope (UKIRT)
Infrared Deep Sky Survey \citep[UKIDSS;][]{Lawrence2007}
and the 
VISTA Variables in the Via Lactea survey \cite[VVV;][]{Minniti2010}.
There is a significant degree of overlap
between the footprints of UKIDSS Galactic Plane Survey \citep[GPS;][]{Lucas2008}
and IPHAS,
and hence by adopting a similar catalogue format
we hope to facilitate scientific applications
which combine both data sets.

A detailed description of each column in our source catalogue
is given in Appendix~\ref{app:columns}.
In the remainder of this section we highlight the main features.

First, we note that each source is uniquely identified by an
IAU-style designation of the form `IPHAS2\ JHHMMSS.ss+DDMMSS.s'
(cf. column \emph{name} in Appendix~\ref{app:columns}),
where `IPHAS2' refers to the present
data release and the remainder of the string
denotes the J2000 ICRS coordinates in sexagesimal format.
For convenience, the coordinates
are also included in decimal degrees
(columns \emph{ra} and \emph{dec})
and in Galactic coordinates
(columns \emph{l} and \emph{b}).
We have also included an internal object identifier string 
of the form `\#run-\#ccd-\#detection'
(e.g. `64738-3-6473'),
which documents the INT exposure number (\#run),
the CCD number (\#ccd),
and the row number in the CASU detection table (\#detection).
These columns are named \emph{rDetectionID},
\emph{iDetectionID}, \emph{haDetectionID}.

Photometry is provided based on the 2\farcs 3-arcsec diameter circular aperture
by default (columns \emph{r}, \emph{i}, \emph{ha}).
The choice of this aperture size as the default 
is based on a trade-off between concerns 
about small number statistics and centroiding errors
for small apertures on one hand,
and diminishing signal-to-noise ratios and source confusion
for large apertures on the other hand.
The user is not restricted to this choice, because
the catalogue also provides magnitudes
using three alternatives:
the peak pixel height 
(columns \emph{rPeakMag}, \emph{iPeakMag}, \emph{haPeakMag}),
the circular 1.2-arcsec-diameter aperture 
(\emph{rAperMag1}, \emph{iAperMag1},
 \emph{haAperMag1}) and
the 3.3-arcsec-diameter aperture 
(\emph{rAperMag3}, \emph{iAperMag3},
 \emph{haAperMag3}).

Each of these magnitude measurements have been
corrected for the flux lost outside of the respective apertures,
using a correction term which is inferred from the
mean shape of the PSF measured locally in the CCD frame.
In the case of a point source,
the four alternative magnitudes are expected
to be consistent with each other to
within the photon noise uncertainty
(which is given in columns \emph{rErr}, \emph{rPeakMagErr},
\emph{rAperMag1Err}, \emph{rAperMag3Err}, etc).
When this is not the case,
it is likely that the source is either
an extended object for which the aperture correction is invalid,
or that the object has been incorrectly measured as a result of
source blending or a rapidly spatially-varying nebulous background.
In \S\ref{sec:qualitycriteria} we will explain that the consistency
of the different-aperture magnitude measurements
can be used as a criterion for selecting stellar objects
with accurate photometry.

The brightness of each object as a function of increasing
aperture size is also used by the CASU pipeline to provide
a discrete star/galaxy\footnote{For consistency with the terminology that is used in the CASU pipeline and the WSA/VSA archives, extended objects are classified as `galaxies'. This class will flag any type of resolved object, however.}/noise classification flag classification flag
(\emph{rClass}, \emph{iClass}, \emph{haClass})
and a continuous stellarness-of-profile statistic
(\emph{rClassStat}, \emph{iClassStat}, \emph{haClassStat}).
For convenience, we have combined
these single-band morphological measures
into band-merged class probabilities and flags
using the merging scheme in use at the WSA\footnote{Explained at
http://surveys.roe.ac.uk/wsa/www/gloss\_m.html \#gpssource\_mergedclass
} (\emph{pStar}, \emph{pGalaxy}, \emph{pNoise},
\emph{mergedClass}, \emph{mergedClassStat}).

Information on the quality of each detection is included 
in a series of additional columns.
We draw attention to three useful flags
which warn about the likely presence of a systematic error:
\begin{enumerate}
\item The \emph{saturated} column is used to flag sources
for which the peak pixel height exceeds 55000 counts,
which is typically the case for stars
brighter than 12-13th magnitude in $r$.
Although the pipeline attempts to extrapolate the brightness of
saturated stars based on the shape of the PSF,
such extrapolation is prone to error,
and should be viewed as indicative rather than as precise measurement
(systematic uncertainties as a function of magnitude
will be discussed in \S\ref{sec:uncertainties}).
\item The \emph{deblend} column is used to flag sources 
which partially overlap with a nearby neighbour.
Although the pipeline applies a deblending procedure
to such objects, the procedure is currently applied separately
in each band, and hence the $(r-i)$ and $(r-\ha)$ colours
may be inaccurate if the deblending proceeded differently in each band.
\item The \emph{brightNeighb} column is used to flag sources which are located
within a radius of 5 arcmin from an object brighter than $V=7$ 
according to the Bright Star Catalogue (BSC; Hoffleit et al. 1991), 
or within 10 arcmin if the neighbour is brighter than $V=4$.
These brightest stars are known to cause systematic errors
and spurious detections as a result of stray light 
and diffraction spikes.
\end{enumerate}
In addition to the above, we also created warning flags for internal bookkeeping.
For example, we flagged detections which fell in the strongly vignetted regions of the focal plane,
which were truncated by CCD edges,
or which were otherwise affected by bad pixels in the detector.
No such detections have had to be included in the catalogue, as 
alternative detections were available in essentially all these situations
thanks to the IPHAS field pair strategy. 
Hence there has been no need to include these internal warning flags
in the published source catalogue.

Finally, we note that basic information on the observing conditions
is included (\emph{fieldID}, \emph{fieldGrade}, \emph{night}, \emph{seeing}).
A table containing more detailed quality control information,
indexed by \emph{fieldID}, is made available on our website.

\subsection{Band-merging the detection tables}

The second step in compiling the source catalogue
is to merge the contemporaneous trios
of $r$, $i$, \ha\ detection tables
into multi-band field catalogues.
This required a position matching procedure 
to link sources between the three bands.
We used the TMATCHN function 
of the STILTS software
for this purpose,
which allows rows from multiple tables
to be matched \citep{Taylor2006}.
In brief, the algorithm
identifies groups of linked detections such that
(i) each detection is located within a specified maximum distance from
one or more members of the group,
(ii) each detection appears in exactly one group, and
(iii) the largest possible groups are preferred
(i.e. preferably containing three detections from all three bands).
The result of the procedure is a band-merged catalogue
in which each row corresponds
to a group of linked $r$, $i$, and \ha\ detections
which satisfy the matching distance criterion
in a pair-wise sense.
Sources for which no counterpart was identified
in all three bands are retained in the catalogue
as single- or double-band detections,
with empty columns for the missing bands.

We employed a maximum matching distance of 1~arcsec,
trading off completeness against reliability.
On the one hand, a matching distance larger than 1~arcsec 
was found to allow too many spurious and unrelated sources 
to be linked. 
On the other, a value smaller than 1~arcsec 
would pose problems for very faint sources 
with large centroiding errors, 
and would occasionally fail near CCD corners,
where the astrometric solution can 
show local systematic errors which exceed 0.5~arcsec.
The position offsets between the $r$ detection and detections in $i$, and/or \ha\
have been included in the catalogue, giving the user the option to tighten them 
further if necessary
(columns \emph{iXi}, \emph{iEta}, \emph{haXi}, \emph{haEta}),
or simply to examine light centre differences.
We note that UKIDSS/GPS adopted 
the same maximum matching distance
for similar reasons \citep{Hambly2008}.

The resulting band-merged catalogues were inspected by eye
as part of our quality control procedures
and were found to be reliable for the vast majority of objects.
We do warn that blended objects
can occasionally fall victim to source confusion
during the band-merging procedure,
which is a complicated problem
that we have not attempted to resolve in this release.
It is important to bear this in mind when appraising stars
of seemingly unusual colours (such as candidate emission line stars).
If blending is flagged, or if the inter-band matching distance
is unusually large, then the probability that the unusual colour is 
spurious due to source confusion is greatly increased.

\subsection{Selecting the primary detections}

We explained earlier that the survey contains repeat observations
of identical sources as a result of field offsetting and overlaps.
Amongst all sources in the magnitude range $13<r<19$,
we find that 65 per cent were detected twice
and 25 per cent were detected three times or more.
Only 9 per cent were detected once.

Since the principal aim of this data release is to provide 
accurate photometry at a single epoch,
we have focused on providing the magnitudes
and coordinates from the \emph{best-available} 
detection of each object -- 
hereafter referred to as the \emph{primary} detection.
Although overlapping fields could have been co-added 
to gain a small improvement in depth, 
we have decided against this for two reasons.
Firstly, combining the information from multiple epochs
would make the photometry of variable stars difficult to interpret.
Secondly, co-adding would cause the image quality to degrade towards the mean,
which is particularly a drawback for crowded fields.

Anyone interested in the alternative detections of a source
-- hereafter called the \emph{secondary} detections --
can nevertheless obtain this information in two ways.
To begin with, whenever a secondary detection was collected 
within 10 minutes of the primary,
we have included the identifier and the photometry
of that secondary detection
in the catalogue for convenience
(columns \emph{sourceID2}, \emph{fieldID2}, 
\emph{r2}, \emph{i2}, \emph{ha2},
\emph{rErr2}, \emph{iErr2}, \emph{haErr2}, \emph{errBits2}).
Second, images not included in the catalogue
are made available on our website.

Primary detections have been selected from all available detections using a so-called \emph{seaming} procedure,
which we adapted from the algorithm developed for the WSA\footnote{http://surveys.roe.ac.uk/wsa/dboverview.html\#merge}.
In brief, the first step is to identify all the duplicate detections
by cross-matching the overlap regions of all field catalogues,
again using a maximum matching distance of 1~arcsec.
The duplicate detections for each unique source are then ranked according to
(i) filter coverage, (ii) quality score,
and (iii) the average seeing of stars in the CCD frame rounded to 0.2~arcsec.
If this ranking scheme reveals multiple `winners' of seemingly identical quality,
then the one that was observed closest to the optical axis of the camera is chosen.

\subsection{Compiling the final source catalogue}

As the final step, the primary detections
selected above were compiled
into the final 99-column source catalogue
that is described in Appendix~\ref{app:columns}.
The original unweeded list of sources naturally included 
a significant number of spurious entries
as a result of the sensitive detection level
that is employed by the CASU pipeline.
We have decided to enforce three basic criteria
which must be met for a candidate source
to be included in the catalogue:
\begin{enumerate}
\item the source must have been detected at S/N$>5$ in at least
one of the bands, i.e. it is required that at least one of
\emph{rErr}, \emph{iErr} or \emph{haErr} is smaller
than 0.2 mag;
\item the shape of the source must not be an obvious
cosmic ray or noise artefact, i.e. we require
either \emph{pStar} or \emph{pGalaxy} to be
greater than 20 per cent;
\item the source must not have been detected
in one of the strongly vignetted corners of the instrument, 
not have had any known bad pixels in the aperture,
and not have been on the edge of one of the CCDs.
\end{enumerate}

A total of 219 million primary detections satisfied
the above criteria and have been included in the catalogue.

Table~\ref{tbl:detections} details the breakdown of these
sources as a function of the bands in which they are captured.
159 million sources are detected in all three filters (73 per cent),
30 million are detected in two filters (14 per cent),
and the remaining 30 million are single-band detections.
Table~\ref{tbl:detections} also presents 
the fraction of `confirmed' objects,
which we define as those sources
which have been detected more than once
(recall that much of the survey area
is observed twice due to the field pair strategy).
We find that the single-band detections tend
to show much lower confirmation rates (32 per cent on average)
than double- and triple-band detections (89 per cent).
This suggests that a significant fraction of
these entries may be spurious detections.

\begin{table}
    \caption{Breakdown of catalogue sources as a function of the band(s)
             in which the object was detected.
             We also show the fraction of `confirmed' sources,
             which we define as those objects
             detected in more than one field
             (usually the field pair partner).}
    \label{tbl:detections}
    \begin{center}
        \begin{tabular}{lrcccc}
        \toprule
        Band(s) & Sources & Confirmed \\
        & [$10^6$] & (\emph{nObs}$>$1) \\
        \midrule
        $r$, $i$, \ha & 159 & 91\% \\
        $r$, $i$ & 25 & 77\% \\
        $i$, \ha & 3 & 73\% \\
        $r$, \ha & 2 & 65\% \\
        $i$ & 15 & 43\% \\
        $r$ & 9 & 27\% \\
        \ha & 6 & 12\% \\
        \midrule
        Total  & 219 & 81\% \\
        \bottomrule
        \end{tabular}
	\end{center}
\end{table}

Not all the single-band detections are spurious, however.
We note that the confirmation rate for $i$-band detections
is markedly better than for $r$ and \ha,
which is likely to be explained by the fact that $i$ is least
affected by interstellar extinction, and so the survey can occasionally pick up
highly-reddened objects in $i$ which are otherwise lost in $r$ and \ha.
Moreover, objects which are intrinsically very red may also be picked
up in $i$ alone, while faint objects with very strong Balmer emission
may appear only in \ha.
Nevertheless, we recommend users not to rely on single-band objects
without inspecting the image data by eye,
or verifying that the object
was detected more than once (\emph{nObs}$>$1).

\section{Discussion}
\label{sec:discussion}

We now offer an overview of the properties of the
catalogue by discussing 
(i) the known caveats,
(ii) the recommended quality criteria,
(iii) the reliability of sources,
(iv) the photometric uncertainties,
and (v) the source densities.

\subsection{Known caveats}

Like any other photometric survey
in which a majority of detected objects are close
to the detection and resolution limits,
our catalogue inevitably contains sources
that are spurious or have been parametrised inaccurately.
In what follows, we highlight the most common
caveats which users of the catalogue might face,
followed by a discussion on how they can be avoided.
These caveats include:
\begin{itemize}
\item \emph{Spurious objects.}
Nebulous sky backgrounds,
saturation artefacts near bright stars,
and cosmic rays
are known to be able to trigger spurious detections.
The majority of these can be removed by requiring
that a detection is made in more than one band
(\emph{nBands}$>$1),
on more than one occasion (\emph{nObs}$>$1),
or by ensuring that the object looks like a perfect
point source (\emph{pStar}$>$0.9).
\item \emph{Source blending and confusion.}
Blended objects are known to be parametrised less accurately
and to be more prone 
to source confusion
during the band-merging procedure.
We remind the reader that such objects are flagged in the catalogue
using the columns \emph{rDeblend}, \emph{iDeblend},
\emph{haDeblend}, and \emph{deblend}.
\item \emph{Low S/N detections.}
The majority of the objects in our catalogue
are faint sources observed near the detection limits,
e.g. 55 per cent of the entries in the catalogue
are fainter than $r > 20$.
The measurements of faint objects
are naturally prone to larger
random and systematic uncertainties:
for example, an inaccurately-subtracted background
will introduce a proportionally larger systematic error
for a faint object.
These objects can be removed by ensuring that an
object is detected at $S/N>10$
and has photometry which is consistent across
the different aperture sizes.
Objects detected at $S/N>10$ in all bands
are flagged in the catalogue using the \emph{a10} column.
\item \emph{Saturation.}
The photometry and astrometry of objects
brighter than the saturation limit of the instrument
is subject to systematic errors.
Such objects can be removed by ensuring that the columns
\emph{rSaturated}, \emph{iSaturated}, \emph{haSaturated}
or \emph{saturated} are set to ``False''.
\end{itemize}
We note that it is not possible, at this stage,
to include an exhaustive list of all the caveats,
because we cannot anticipate all forms of use of the data.
For this reason, we recommend users to read
the FAQ section on our website,
which will be updated as user experience accumulates.

\subsection{Recommended quality criteria}
\label{sec:qualitycriteria}

Many applications will require
a combination of quality criteria
to be used
to avoid the issues identified above.
The choice of criteria will always tension 
completeness against reliability,
i.e. the fraction of spurious sources.
To aid users we have listed two sets of
recommended quality criteria 
in Tables~\ref{tab:a10} and \ref{tab:a10point}.

First, Table~\ref{tab:a10} specifies
a set of minimum quality criteria
which should benefit most applications
which desire accurate colours
as well as completeness down to $\sim$19th magnitude.
The listed criteria are designed
to ensure that each band offers photometry
at S/N\ $>10$ that is self-consistent
across different aperture sizes.
A total of 86 million sources out of 219 million 
(39 per cent) satisfy all the criteria listed in Table~\ref{tab:a10}
and are hereafter referred to as `a10'
(short for ``all-band $10\sigma$'').
For convenience, the catalogue contains a boolean column
named \emph{a10} that flags these objects.

\begin{table*}
\vspace{2.5cm}
\caption{Recommended quality criteria 
for selecting objects with accurate colours.
These criteria serve to identify objects detected 
at S/N$\,>\,$10 in all three bands without being saturated,
and with the added requirement that the photometric measurements
need to be consistent across different aperture sizes.
The 86~million objects which satisfy these criteria
have been flagged in the catalogue using the column named \emph{a10}
(short for ``all-band 10$\sigma$'').}
\begin{tabular}{p{8cm}lp{6cm}}
  \hline
  Quality criterion & Rows passed & Description \\
  \hline
  $rErr\,<\,0.1$ {\sc and} $iErr\,<\,0.1$ {\sc and} $haErr\,<\,0.1$ &
  109 million (50\%) &
  Require the photon noise to be less than 
  0.1 mag in all bands (i.e. S/N$>$10).
  This implicitly requires a detection in all three bands.  \\
  {\sc not} \emph{saturated} &
  158 million (72\%) &
  The brightness must not exceed the nominal saturation limits.
  \\
  $|r- rAperMag1| 
  < 3\sqrt{rErr^2 + rAperMag1Err^2} 
  + 0.03$ &
  176 million (80\%) &
  Require the $r$ magnitude measured 
  in the default 2\farcs 3-diameter aperture
  to be consistent with the measurement 
  made in the smaller 1\farcs 2 aperture,
  albeit tolerating a 0.03 mag systematic error.
  This will reject sources for which the background
  subtraction or the deblending procedure
  was not performed reliably. \\
  $|i- iAperMag1| 
  < 3\sqrt{iErr^2 + iAperMag1Err^2} 
  + 0.03$ &
  183 million (84\%) &
  Same as above for $i$. \\
  $|ha- haAperMag1| < 
  3\sqrt{haErr^2 + haAperMag1Err^2} 
  + 0.03$ &
  158 million (72\%) &
  Same as above for \ha. \\
  \hline
  All of the above (flagged as {\bf\emph{a10}}) &
  86 million (39\%) & \\
  \hline
\end{tabular}
\label{tab:a10}
\vspace{2.5cm}
\caption{Additional quality criteria which are recommended
for applications which require objects to be
single, unconfused point sources with accurate colours.
The 59 million sources which satisfy these criteria 
have been flagged using the column named \emph{a10point}.}
\begin{tabular}{p{8cm}lp{6cm}}
  \hline
  Quality criterion & Rows passed & Description \\
  \hline
   \emph{a10} &
   86 million (39\%) &
   The object must satisfy the criteria for accurate colours listed in Table~\ref{tab:a10}. \\
   
   $pStar > 0.9$ &
   145 million (66\%) &
   The object must appear as a perfect point source,
   as inferred from comparing its PSF
   with the average PSF measured in the same CCD. \\
   
   {\sc not} \emph{deblend} &
   177 million (81\%) &
   The source must appear as a single, unconfused object. \\
   
   {\sc not} \emph{brightNeighb} &
   216 million (99\%) &
   There is no star brighter than $V < 4$ within 10\arcmin, 
   or brighter than $V < 7$ within 5\arcmin.
   Such very bright stars cause scattered light and diffraction spikes,
   which may add systematic errors to the photometry
   or even trigger spurious detections. \\  
  \hline
  
  All of the above (flagged as {\bf\emph{a10point}}) &
  59 million (27\%) & \\
  \hline
\end{tabular}
\label{tab:a10point}
\vspace{1cm}
\end{table*}

\begin{figure}
    \includegraphics[width=0.45\textwidth]{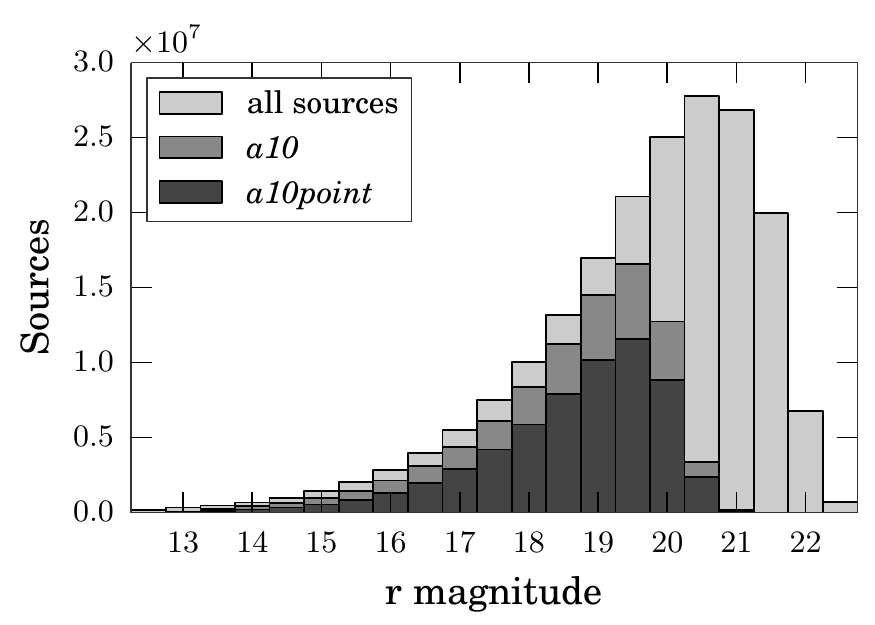} 
    \caption{r-band magnitude distribution
    for all objects in the catalogue (light grey).
    Overlaid we also show the distribution for objects
    detected at S/N $>$ 10 in all bands
    selected according to the quality criteria 
    described in Table~\ref{tab:a10} (\emph{a10}, grey),
    and for the set of unconfused 10$\sigma$ point source detections
    described in Table~\ref{tab:a10point} (\emph{a10point}, dark grey).
    The distributions for $i$ and \ha\
    look identical, apart from being shifted
    by about 1 and 0.5 mag towards brighter magnitudes,
    respectively.}
    \label{fig:magdist}
\end{figure}

For applications which require
a higher standard of accuracy at the expense of completeness,
a set of additional quality criteria
are suggested in Table~\ref{tab:a10point}.
These criteria are designed to ensure that
(i) the object appeared as a perfect point source in all bands,
(ii) the object was not blended with a nearby neighbour,
and (iii) the object was not located near a very bright star.
59 million sources (27 per cent) satisfy
these stricter criteria 
and are hereafter referred to as `a10point'.
Again, the catalogue contains a boolean column
named \emph{a10point} which flags these objects.

Fig.~\ref{fig:magdist} compares the $r$-band magnitude
distributions for objects tagged \emph{a10} and \emph{a10point}
against the unfiltered catalogue.
we find that 81 per cent of sources 
in the magnitude range $13 < r < 19$
are flagged \emph{a10},
and 54 per cent are flagged \emph{a10point}.
We will show below that the
\emph{a10point} category is least complete
at low Galactic longitudes,
where source blending can affect
up to a quarter of the objects.

It is easy to see how the quality criteria
may be adapted to be more tolerant.
For example, by raising the allowed photometric uncertainties
from 0.1~mag to 0.2~mag in Table~\ref{tab:a10},
42 million candidate sources would be added to the 109 
million satisfying the tighter error bound.
Our choice to adopt 0.1~mag as the cut-off uncertainty
for the \emph{a10} category is a pragmatic trade-off
which we found to suit many science applications,
but users are encouraged 
to revise the quality criteria according to their needs.

\subsection{Source reliability}
\label{sec:reliability}

Having obtained a set of quality criteria,
we now assess how these improve the data quality.
In this section we start by discussing
the \emph{reliability} of the survey,
which we define as the fraction of sources
which are genuine astrophysical objects.
To estimate the reliability, we cross-matched IPHAS
against SDSS DR7\footnote{SDSS DR7 was chosen
due to an apparent change in 
the catalogue preparation for DR8 and DR9,
which resulted in fewer genuine stars being
detected near bright ($r\,<\,12$) sources,
which in turn gave an unduly pessimistic view
on our reliability.},
which is a catalogue of similar resolution and depth.
By computing the fraction of IPHAS sources which were
also independently detected by SDSS,
we can obtain a good lower limit on the reliability.

The comparison was carried out across three cone-shaped regions,
each covering 1 deg$^2$,
using a strict cross-matching distance of 0\farcs 5.
We considered four subsets of the IPHAS catalogue:
\begin{enumerate}
\item the entire (unfiltered) catalogue;
\item objects detected in all three bands;
\item \emph{a10} sources;
\item \emph{a10point} sources.
\end{enumerate}

\begin{table}
    \caption{Fractions of sources in our catalogue
             which are also found in SDSS DR7
             within a cross-matching radius of 0\farcs 5.
             These percentages can be interpreted
             as lower limits
             for the reliability of our sources,
             i.e. the probability that an entry in our
             catalogue is a genuine astrophysical source
             rather than an instrumental artefact.
             The fractions were computed
             in three 1 deg$^2$ regions
             where both surveys
             show contiguous overlap.
             The reliability is shown
             without quality criteria applied (``all''),
             using only objects detected in all three bands ($nBands=3$),
             using objects detected at 10$\sigma$
             in all three bands (\emph{a10}),
             and finally using objects
             classified as reliable 10$\sigma$ point sources (\emph{a10point}).
             }
    \label{tbl:reliability}
    \begin{center}
        \begin{tabular}{lcccccc}
        \toprule
        Region ($\ell$, $b$) & all & $nBands=3$ & \emph{a10} & \emph{a10point} \\
        \midrule
        (149.39, 4.06) & 93.0\% & 98.8\% & 99.7\% & 99.8\% \\
        (186.59, -2.50) & 90.0\% & 98.8\% & 99.4\% & 99.6\% \\
        (202.70, -1.75) & 89.4\% & 98.5\% &99.4\% & 99.7\% \\ \midrule
        mean & 90.8\% & 98.7\% & 99.5\% & 99.7\% \\
        \bottomrule
        \end{tabular}
    \end{center}
\end{table}

The results for each of the regions and subsets
are presented in Table~\ref{tbl:reliability}.
We find that our catalogue shows a good
baseline reliability (90.8 per cent),
which is improved markedly by the simple requirement
that a source needs to be detected in all three bands
(98.7 per cent). 
Reliabilities of 99.5 and 99.7 per cent are achieved
using the \emph{a10} and \emph{a10point} quality flags.

To understand the nature of the small fraction
of \emph{a10point}-flagged sources
which appear to be unreliable in this test,
we investigated the data by eye.
We found these `unreliable' objects to be located
in the vicinity of moderately bright stars ($r\,\lesssim\,12$),
tracing out an area which is affected by saturation spikes or
scattered light in the SDSS images.
In all cases we found these remaining objects 
to appear as genuine stars in both the IPHAS and SDSS images.
It is hence likely that the reliability
of the \emph{a10point} class is close to 100 per cent.

\subsection{Random and systematic uncertainties}
\label{sec:uncertainties}

\begin{figure}
    \includegraphics[width=0.45\textwidth]{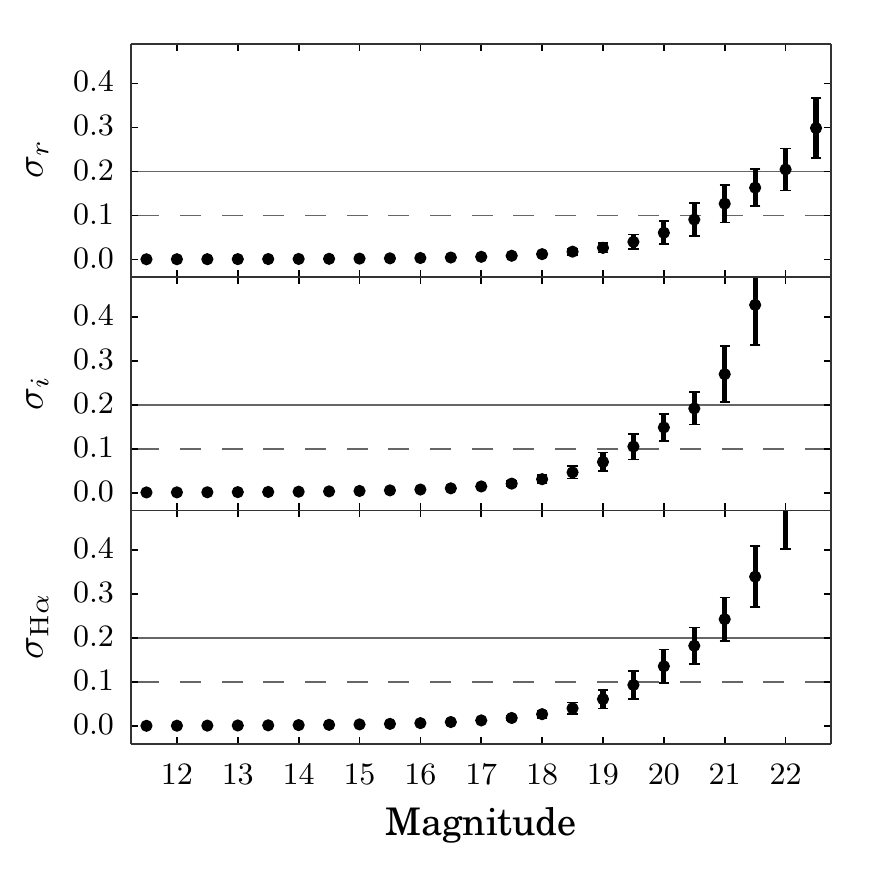} 
    \caption{Mean photometric uncertainties
             for $r$ (top), $i$ (middle) and \ha\ (bottom).
             Data points shown are the average values of
             columns \emph{rErr}, \emph{iErr} and \emph{haErr}
             in the catalogue,
             and the error-bars show the standard deviations.
             The dashed and solid lines indicate 
             the 10$\sigma$ and 5$\sigma$ limits, respectively.
             These statistics show the average level of the Poissonian
             photon noise and do not include systematic uncertainties.}
    \label{fig:uncertainties}
\end{figure}

Fig.~\ref{fig:uncertainties} shows the mean photometric
uncertainties (\emph{rErr}, \emph{iErr}, \emph{haErr})
as a function of magnitude.
We find the typical uncertainty to reach 0.1~mag near $r=$20.5 
and $i$,\ha$=19.5$.
We note that the fainter depth in $r$ is compensated
by the fact that most stars have brighter magnitudes in $i$ and \ha;
the average colours in the catalogue are
$\overline{(r-i)}=1.06\pm0.12$ and $\overline{(r-\ha)}=0.44\pm0.03$.
We warn that the statistics shown in Fig.~\ref{fig:uncertainties}
are the random errors based on the expected Poissonian photon noise.
Systematics, such as calibration and deblending errors,
are not included.

\begin{figure*}
	\vspace{1cm}
    \begin{minipage}[b]{\linewidth}
        \includegraphics[width=0.5\textwidth]{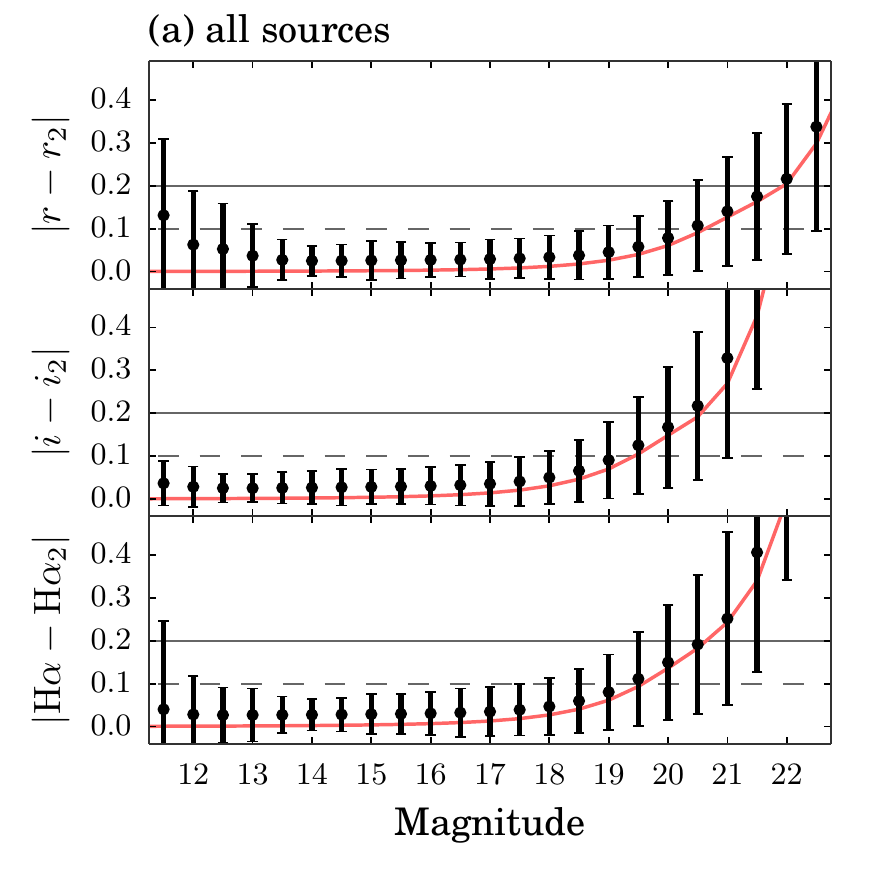} 
        \includegraphics[width=0.5\textwidth]{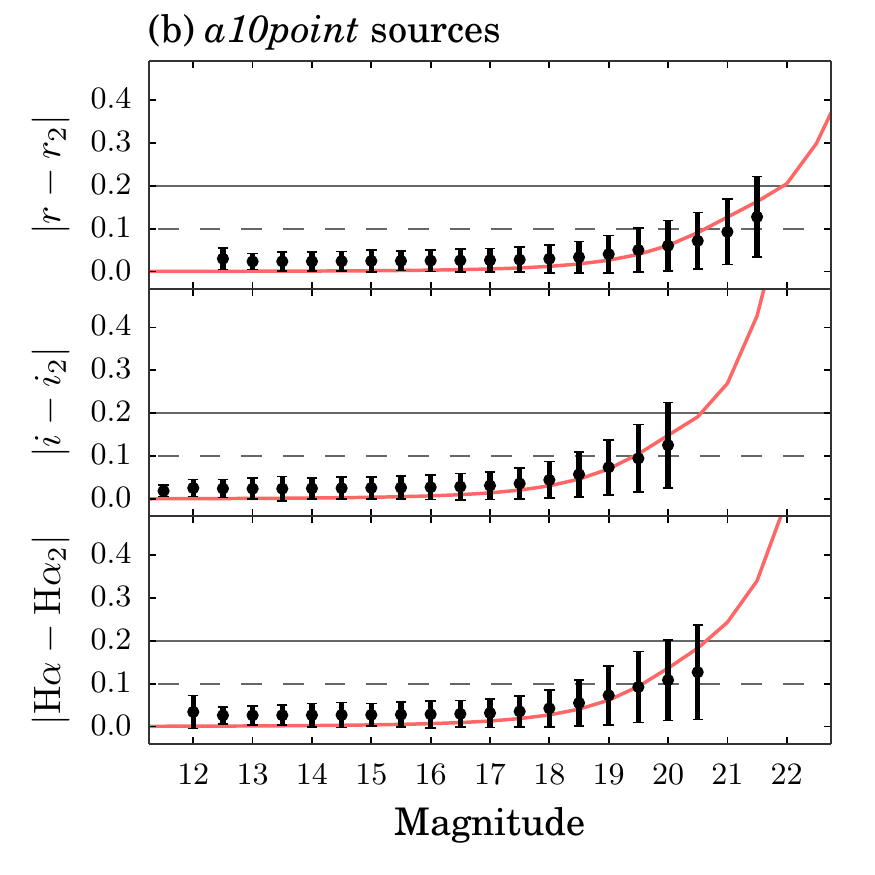}
    \end{minipage}
    \caption{Photometric repeatability as a function of magnitude
         for all sources in the catalogue (panel a)
         and for the \emph{a10point} sources alone (panel b).
         Black dots show the mean absolute residuals
         between the primary and the secondary detections.
         The error-bars show the standard deviations.
         The red trend line shows the average Poissonian uncertainties
         we derived in Fig.~\ref{fig:uncertainties}.
         We find that the \emph{a10point} quality criteria are successful
         at removing objects with large residuals.}
    \label{fig:pairmag}

    \vspace{1.5cm}
    \includegraphics[width=\textwidth]{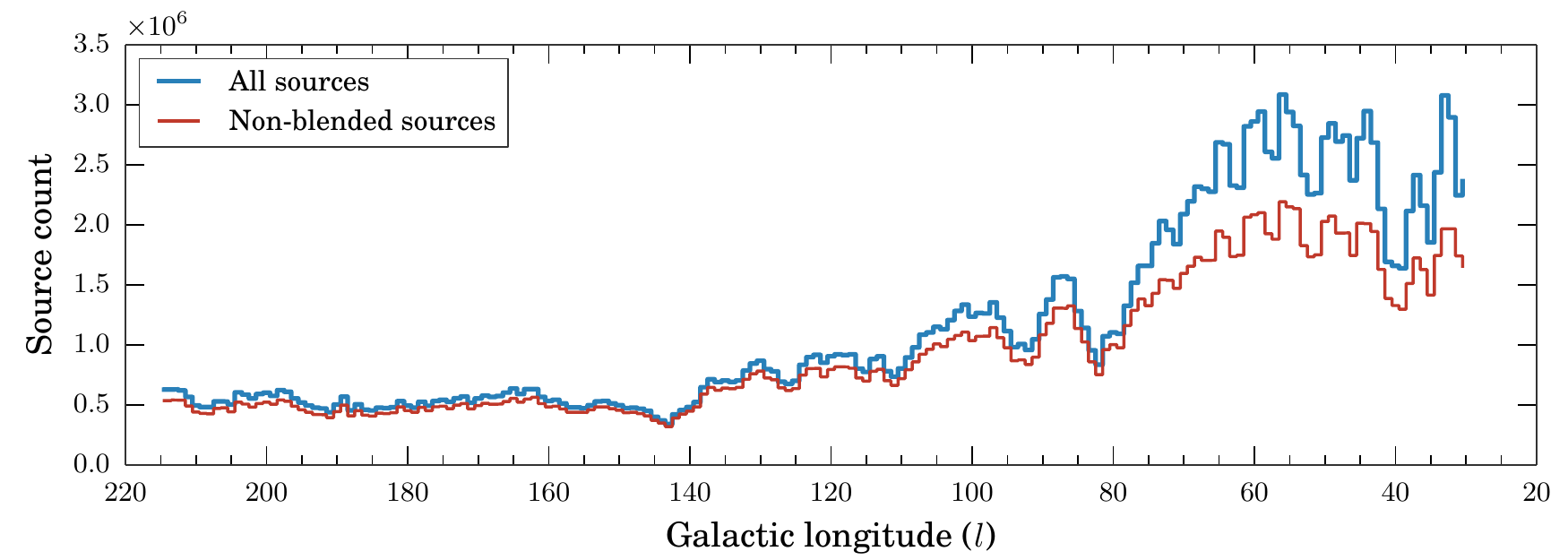} 
    \caption{Number of entries in the IPHAS DR2 source catalogue
    as a function of Galactic longitude.
    The upper blue line shows the number of sources
    counted in 1\degr-wide longitude bins.
    The lower red line uses the same binning
    but includes only those sources 
    for which the \emph{deblend} flag is {\sc false}, 
    i.e. unconfused sources for which the CASU pipeline
    did not have to apply a deblending procedure.
    In both cases we counted only those sources
    in the latitude range $-5\degr<b<+5\degr$,
    such that one may obtain a rough guide to source density
    by dividing the counts by 10~deg$^2$.
    The global variations in the source counts
    traces the structure of the Galaxy
    and the distribution of foreground extinction,
    but is also affected by instrumental effects
	such as variations in the observed depth
	and completeness (see text).
   }
    \label{fig:sourcecount}
\end{figure*}

To appraise the level at which our photometry is affected by such systematics,
we can exploit the secondary measurements which are present in the catalogue
(i.e. \emph{r} vs \emph{r2}, \emph{i} vs \emph{i2}, \emph{ha} vs \emph{ha2}).
In Fig.~\ref{fig:pairmag}a we show the mean absolute residuals between
these primary and secondary magnitudes as a function of magnitude (black dots).
We also plot the Poissonian uncertainties for comparison (solid red line).
We find the mean residual and standard deviation to be $0.03\pm0.04$~mag
across the magnitude ranges 13 to 18 ($r$)
and 12 to 17 ($i$, \ha),
which is consistent with the accuracy of the calibration.
Stars fainter than this range appear to be dominated by photon noise (red line),
while stars at the bright end appear to suffer from large systematic
errors due to saturation effects.

In Fig.~\ref{fig:pairmag}b we show 
a similar comparison between the primary and secondary detections,
but this time we have only included sources which are flagged
as \emph{a10point} in the catalogue.
We do not observe an improvement in the average residuals as a function of magnitude,
but the number of outliers has decreased markedly
(evidenced by the shorter error bars which denote the standard deviation of the absolute residuals).
We conclude that the \emph{a10point} quality criteria are effective
at reducing the level of the systematic errors,
while also removing the inaccurate data at the bright and faint end.

\subsection{Source counts and blending}
\label{sec:densities}

Fig.~\ref{fig:sourcecount} shows the number of sources
in the catalogue counted in 1\degr-wide strips
as a function of Galactic longitude (thick blue line).	
Unsurprisingly, we find the number of sources
to increase towards the Galactic centre.
For example, the average source density near $\ell\simeq 30\degr$
is roughly 300,000 objects per square degree,
which is six times more than the density
found near $\ell\simeq 180\degr$.
In addition to the global trend,
variations are also apparent on smaller scales.
For example, we find a significant drop near the constellations 
of Aquila ($\ell\simeq40\degr$) and Cygnus ($\ell\simeq80\degr$ and $\ell\simeq90\degr$),
which are regions known to be affected
by high levels of foreground extinction
\citep[the extremities of `the Great Rift', e.g.][]{BokBok}.
However, we warn that the source counts shown 
have not been corrected for field pairs that have yet
to be released or for variations in the depth across the included fields.
For example, the dip near $\ell\simeq140\degr$
is an artificial feature caused by gaps
in the footprint coverage (seen in Fig.~\ref{fig:footprint}).

In a forthcoming paper, properly-calibrated detailed maps of stellar density
of the northern Galactic Plane will be presented (Farnhill
et al., in preparation).  This will incorporate completeness
corrections based on the statistics of artificial source
recovery.  Such maps are of interest as tests of Galactic models.

Fig.~\ref{fig:sourcecount} also shows the number counts
for non-blended sources (thin red line).
These are sources for which the \emph{deblend} flag is {\sc false},
i.e. sources for which the CASU pipeline did not have to apply 
a deblending procedure to separate the flux
originating from two or more overlapping objects.
This provides some insight into how the fraction of 
blend-affected sources correlates with the local source 
density.  In headline numbers, only 11 per cent of the sources are blended
at $\ell>90\degr$, whereas 24 per cent are blended at $\ell<90\degr$.

Finally, we warn that blended objects
are more likely than unblended objects to have fallen victim
to source confusion during the band-merging procedure.  It is
important to bear this in mind when appraising stars of seemingly
unusual colours (such as candidate emission line stars) -- if 
blending is flagged, the probability that the unusual colour is 
spurious is greatly increased.

\section{Demonstration}
\label{sec:demonstration}

We conclude this paper by demonstrating how the unique
$(r-\ha,\ r-i)$ colour-colour diagram offered by this catalogue
can readily be used to
(i) characterise the extinction regime along different sightlines, and
(ii) identify \ha\ emission-line objects.

\subsection{Colour-colour and colour-magnitude diagrams}

The survey's unique $(r-\ha)$ colour,
when combined with $(r-i)$,
has been shown to provide simultaneous constraints 
on intrinsic stellar colour and interstellar extinction \citep{Drew2008}. 
Put differently, the main sequence in the $(r-\ha,\ r-i)$ diagram
runs in a direction that is at a large angle relative to the reddening vector,
because the ($r-\ha$) colour tends to act
as a coarse proxy for spectral type
and is less sensitive to reddening than $(r-i)$.
As a result, the distribution of a stellar population
in the IPHAS colour-colour diagram
can offer a handle on the properties of the population
and the extinction along a line of sight.

\begin{figure*}
    \begin{minipage}[b]{\linewidth}
        \includegraphics[width=0.5\textwidth]{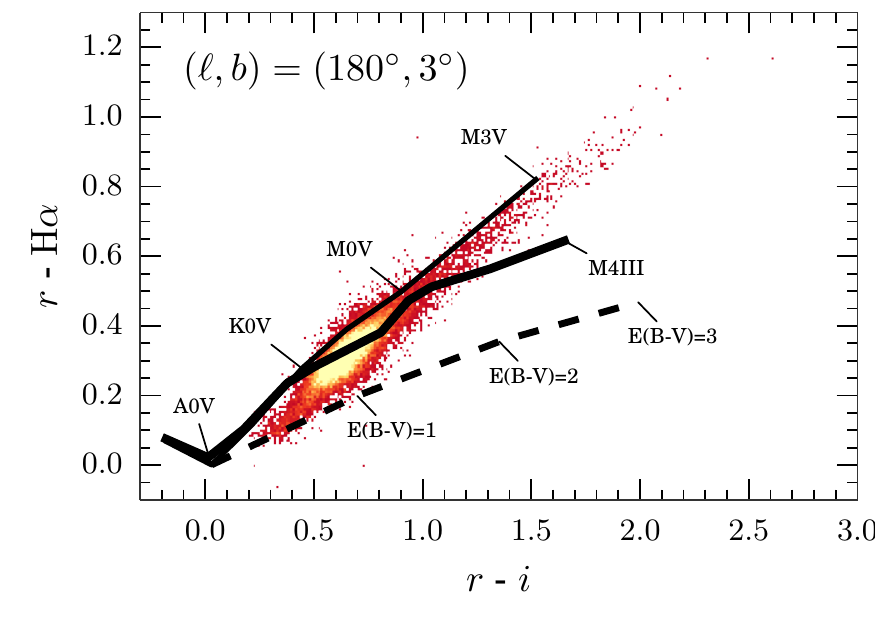} 
        \includegraphics[width=0.5\textwidth]{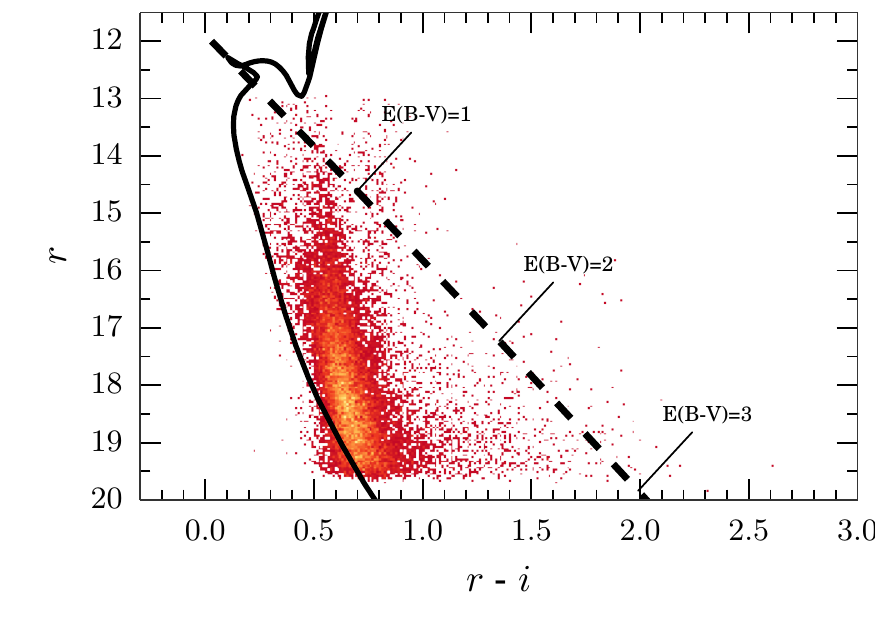}
    \end{minipage}
    \caption{Colour-colour and colour-magnitude diagrams
    (left and right panel)
    showing sources flagged as \emph{a10point}
    located in an area of 1~deg$^2$
    centred near the Galactic anti-centre 
    at $(l,b)=(180\degr,+3\degr)$.
    The diagrams are plotted as 2D-histograms
    which show the density of objects
    in bins of 0.01-by-0.01~mag;
    bins containing 1 to 10 objects are coloured red,
    while bins with more than 20 objects are yellow.
    The left panel is annotated with
    the position of the main sequence (thin solid line),
    giant stars (thick solid line)
    and the reddening track for an A0V-type star (dashed line). 
    The right panel shows the unreddened 1~Gyr isochrone
    from the models by \citet[][solid line]{Bressan2012}
    along with the reddening vector for an A0V-type (dashed line),
    both placed at an arbitrary distance of 2~kpc.
    This is one of the least reddened sightlines
    in the survey
    and hence the observed stellar population appears to be dominated 
    by lowly reddened main sequence stars (see text).}
    \label{fig:l180}
    \begin{minipage}[b]{\linewidth}
        \includegraphics[width=0.5\textwidth]{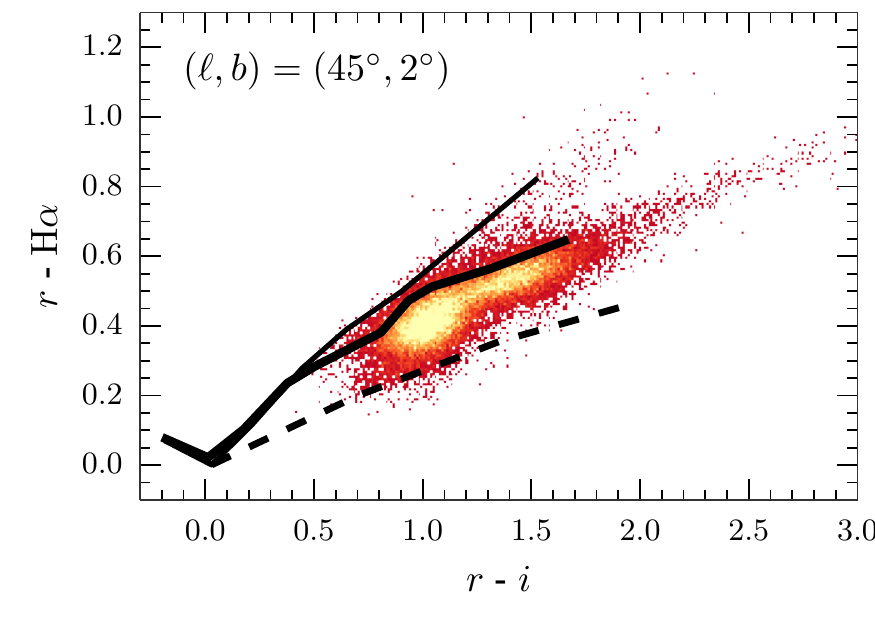}
        \includegraphics[width=0.5\textwidth]{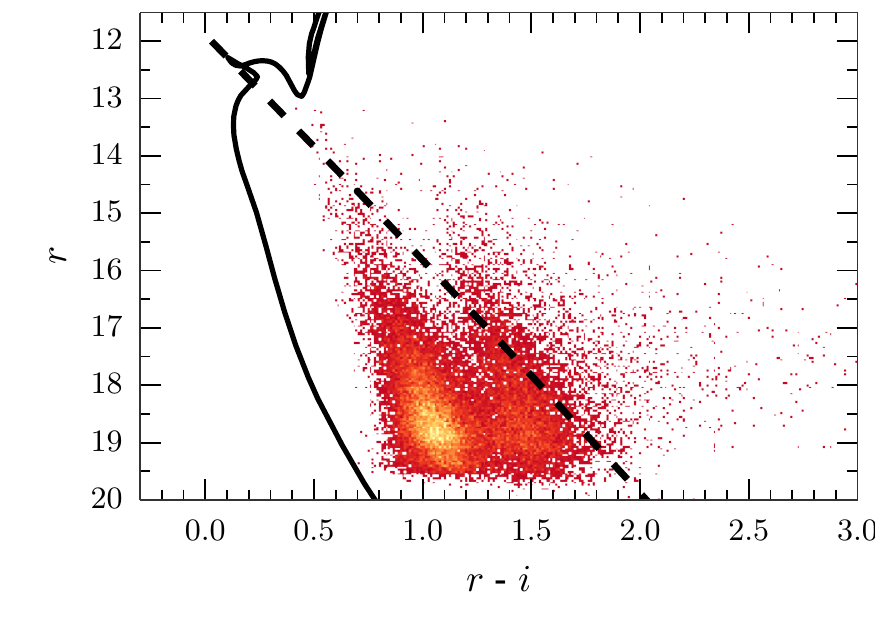}
    \end{minipage}
    \caption{Same as above for $(l,b)=(45\degr,+2\degr)$,
    which is one of the highest-density sightlines in the survey,
    revealing two groups of stars in colour-magnitude space.}
    \label{fig:l45}
    \begin{minipage}[b]{\linewidth}
        \includegraphics[width=0.5\textwidth]{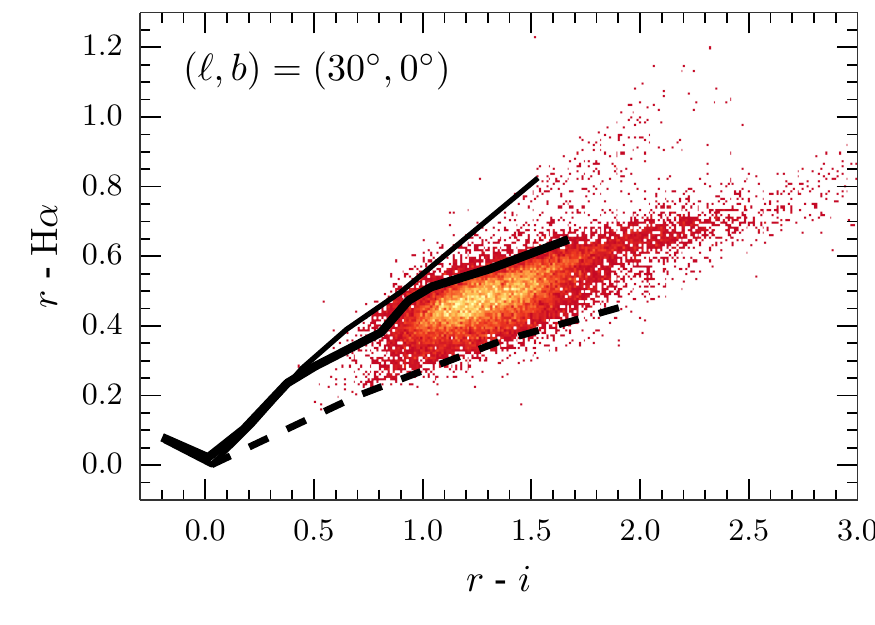}
        \includegraphics[width=0.5\textwidth]{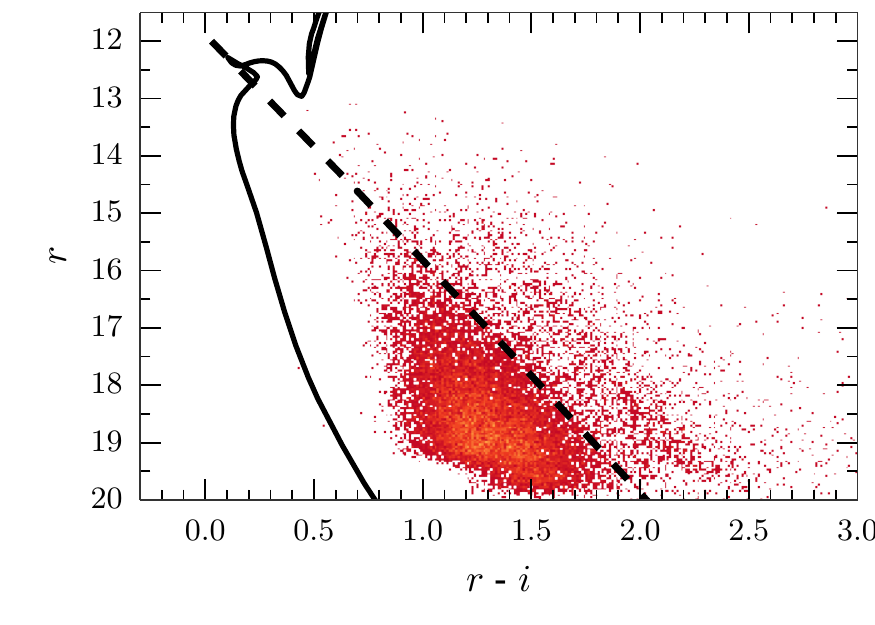} 
    \end{minipage}
    \caption{Same as above for $(l,b)=(30\degr,0\degr)$,
    showing one of the most reddened sightlines in the survey.
    }
    \label{fig:l30}
\end{figure*}

This is demonstrated 
in Fig.~\ref{fig:l180}, \ref{fig:l45} \& \ref{fig:l30},
where we present three sets of IPHAS colour/magnitude diagrams
towards three distinct sightlines
located at Galactic longitudes 
180\degr, 45\degr\ and 30\degr, respectively,
which were chosen because they show stellar populations
with different characteristics.
Each figure contains all the sources
flagged as \emph{a10point} within a region of one square degree 
centred on the coordinates indicated in the diagram
(i.e. within a radius of 0\fdg 564 from the indicated sightline).
For clarity, we have imposed the additional criterion
that the photometric uncertainties
must be smaller than 0.05 mag in each band,
corresponding to a cut-off near 19th magnitude.

Each of the diagrams reveals a well-defined locus,
which helps to further demonstrate the health of the catalogue
and the global calibration
for investigating stellar populations across wide areas.
We have annotated the colour-colour diagrams
by showing the position 
of the unreddened main sequence (thin solid line),
the unreddened giant branch (thick solid line),
and the reddening track for an A0V-type star (dashed line)
-- all three are based on the \cite{Pickles1998} library 
of empirical spectra
synthesised into the Vega-based IPHAS system by \cite{Drew2005}.
In the colour-magnitude diagrams we only show the reddening vector
together with the unreddened 1~Gyr isochrone due to \cite{Bressan2012},
which is made available for the IPHAS system through the
on-line tool hosted at the Observatory of Padova
(http://stev.oapd.inaf.it/cmd).
The isochrone and reddening vector have been placed
at an arbitrary distance of 2~kpc.

Each of the sightlines reveals a stellar population
with distinct characteristics.
Towards the Galactic anti-centre 
at $\ell=180\degr$ (Fig.~\ref{fig:l180})
we find a population dominated by lowly-reddened main sequence stars.
This is consistent with the estimated total sightline extinction 
of $E(B-V)=0.49$ given by \cite{Schlegel1998},
and applying the 14 per cent reduction recommended by \cite{Schlafly2011}.
Looking in more detail we can see
that the stellar locus is narrower for M-type dwarfs
than for earlier types:
we do not observe M dwarfs experiencing 
the strongest reddening possible for this sightline.
This implies that extinction is still increasing 
at distances of $\sim$1-2~kpc,
where M dwarfs become too faint to be contained in the IPHAS catalogue.
It is also clear that there are no unreddened stars
earlier than $\sim$K0 visible;
such stars would be saturated if within a few hundred parsecs.
This therefore suggests 
that there is a measurable increase in extinction locally.
We also note a relative absence of late type giants which,
due to the relative brevity 
of the corresponding phase of stellar evolution,
would only account for a small proportion of this more nearly \emph{volume-limited} sample seen in the Anticentre direction.
 
In contrast, lines of sight passing into the first Galactic quadrant
yield samples that are more commonly \emph{magnitude-limited} instead.
For example, at $\ell=45\degr$ (Fig.~\ref{fig:l45}),
there is  a wealth of reddened objects and late type giants.
In the colour-magnitude diagram, it is clear
that the stars are split into two distinct groups,
with one significantly redder than the other.
The bluer group is composed of main sequence stars,
with the slope of this group in the colour-magnitude diagram
attributable to the significantly increasing extinction.
Meanwhile the redder group is principally composed of red giant stars \citep[see][]{Wright2008}.
As these stars are intrinsically brighter, they will be substantially 
further away than their main sequence counterparts at the same 
apparent magnitude.  Given that extinction continues to increase 
with distance, along this sightline, 
the red giants we observe will be subject to appreciably more reddening
than the main sequence stars, pushing them to $(r - i) \sim 1.5$.

Finally, in one of our lowest-longitude sightlines at $\ell=30\degr$,
we find a very high number of extremely reddened giants
in addition to an unreddened population
of foreground dwarfs.
In contrast to the sightline at $\ell=45\degr$,
there is no clear group of giant stars visible
in the colour-magnitude diagram of Fig.~\ref{fig:l30},
although the red clump stars are manifest as a track of slight 
over-density sitting roughly 0.4~mag redder than the A0V reddening track.
At $(l,b)=(45\degr, +2\degr)$ the giant stars observed
exhibit a relatively narrow range of reddenings
as they lie beyond most of the Galactic dust column.
At $(l,b)=(30\degr, 0\degr)$ this is not the case:
even at the substantial distances
at which we can observe reddened giant stars,
extinction is continuing to rise within the Galactic mid-plane.
It is also apparent that the $(r-i)$ width of both the M dwarfs
and early A dwarfs is greater than that in Fig.~\ref{fig:l45}.
This is indicative of a steeper rise in reddening,
both within several hundred parsecs (M dwarfs)
and within a few kpc (early A dwarfs).

These are just descriptive vignettes of the information obtainable
from IPHAS colour-colour and colour-magnitude plots.
A more rigorous quantitative analysis of the IPHAS catalogue
can be undertaken to estimate both the stellar density distribution 
in the Milky Way \citep{Sale2010} and to create detailed three-dimensional 
maps of the extinction across several kpc \citep{Sale2009,Sale2012}.
A 3-D extinction map based on the DR2 catalogue
is being released in a separate paper \citep{Sale2014}.

\subsection{Identifying \ha\ emission-line objects}

An aim of IPHAS is to enable the discovery 
of new fainter emission-line objects
across the Galactic Plane.
The survey-wide identification and analysis 
of emission-line objects is beyond the scope
of the present work and will be the focus
of a forthcoming paper (Barentsen et al, in preparation).
In this section we merely aim to demonstrate
a use of the catalogue for this purpose.

An initial list of candidate \ha-emitters
based on the first IPHAS data release was previously
presented by \cite{Witham2008}. 
Because no global calibration was available
at the time, \citeauthor{Witham2008} employed 
a sigma-clipping technique to select objects with
large, outlying $(r-\ha)$ colours.
In contrast, the new catalogue
allows objects to be picked out
from the $(r-\ha,\ r-i)$ colour-colour diagram
using model-based colour criteria
rather than an adaptive statistical procedure.
In what follows we demonstrate this new capability 
by selecting candidate emission-line objects
towards a small region in the sky.

The target of our demonstration is Sh 2-82:
a 5~arcmin-wide H{\sc ii} region located near $(l,b)=(53.55\degr, 0.00\degr)$
in the constellation of Sagitta.
Nicknamed by amateur astronomers as the `Little Cocoon Nebula',
Sh 2-82 is ionised by 
the $\sim$10th magnitude star HD\,231616
with spectral type B0V/III
\citep{Georgelin1973,Mayer1973,Hunter1990}.
This ionising star has been placed 
at a likely distance of 1.5-1.7 kpc
based on its photometric parallax
\citep{Mayer1973,Lahulla1985,Hunter1990}.

\begin{figure*}
    \begin{minipage}[b]{0.8\linewidth}
        \includegraphics[width=\textwidth]{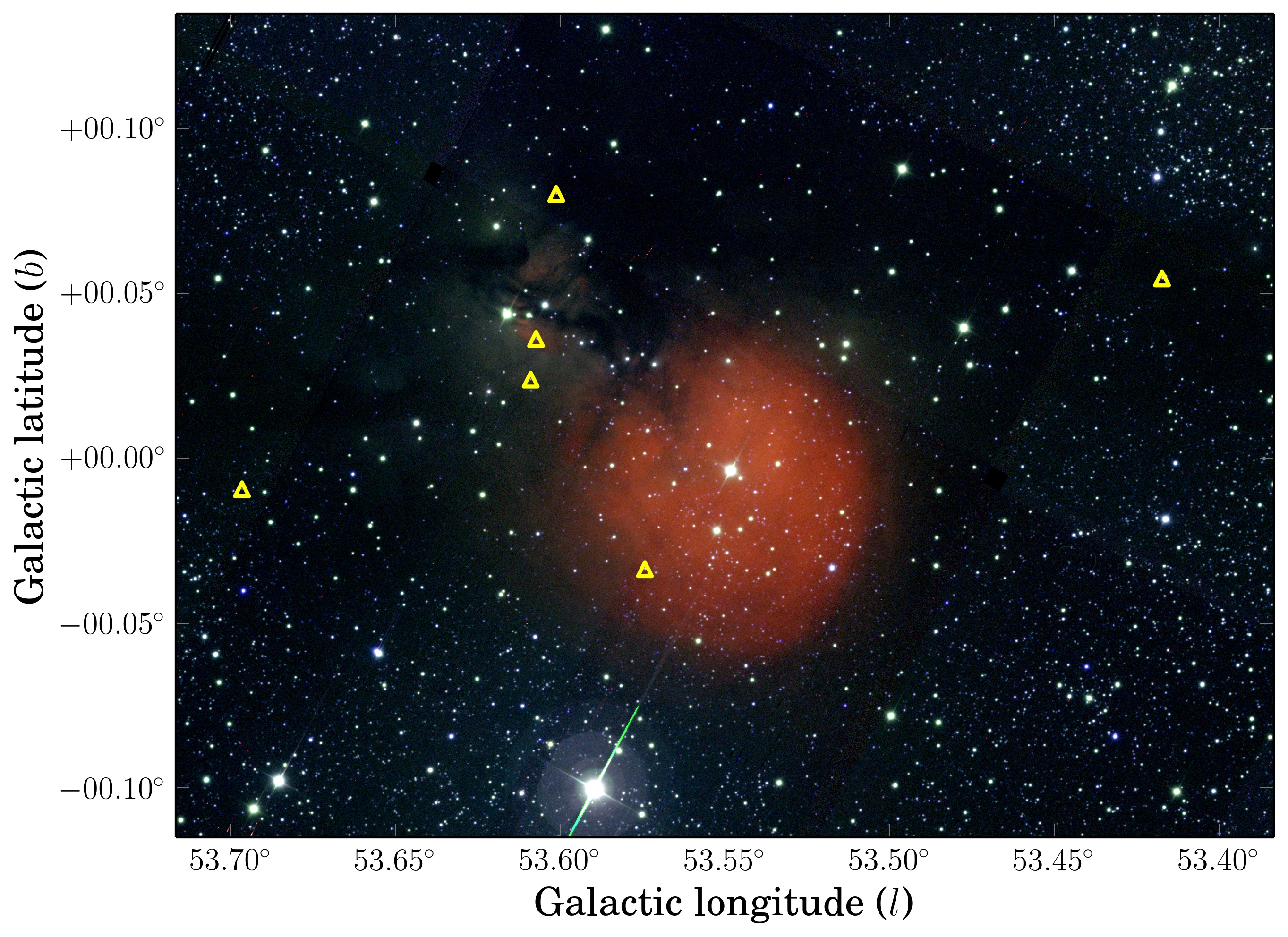} 
    \end{minipage}
\caption{IPHAS image mosaic of H{\sc ii} region Sh 2-82,
composed of \ha\ (red channel), $r$ (green channel) and $i$ (blue channel). Yellow triangles show the position of candidate \ha-emitters
which have been selected from the colour-colour diagram
in Fig.~\ref{fig:emitters}. 
}
\label{fig:mosaic_iphas}
    \begin{minipage}[b]{0.8\linewidth}
        \includegraphics[width=\textwidth]{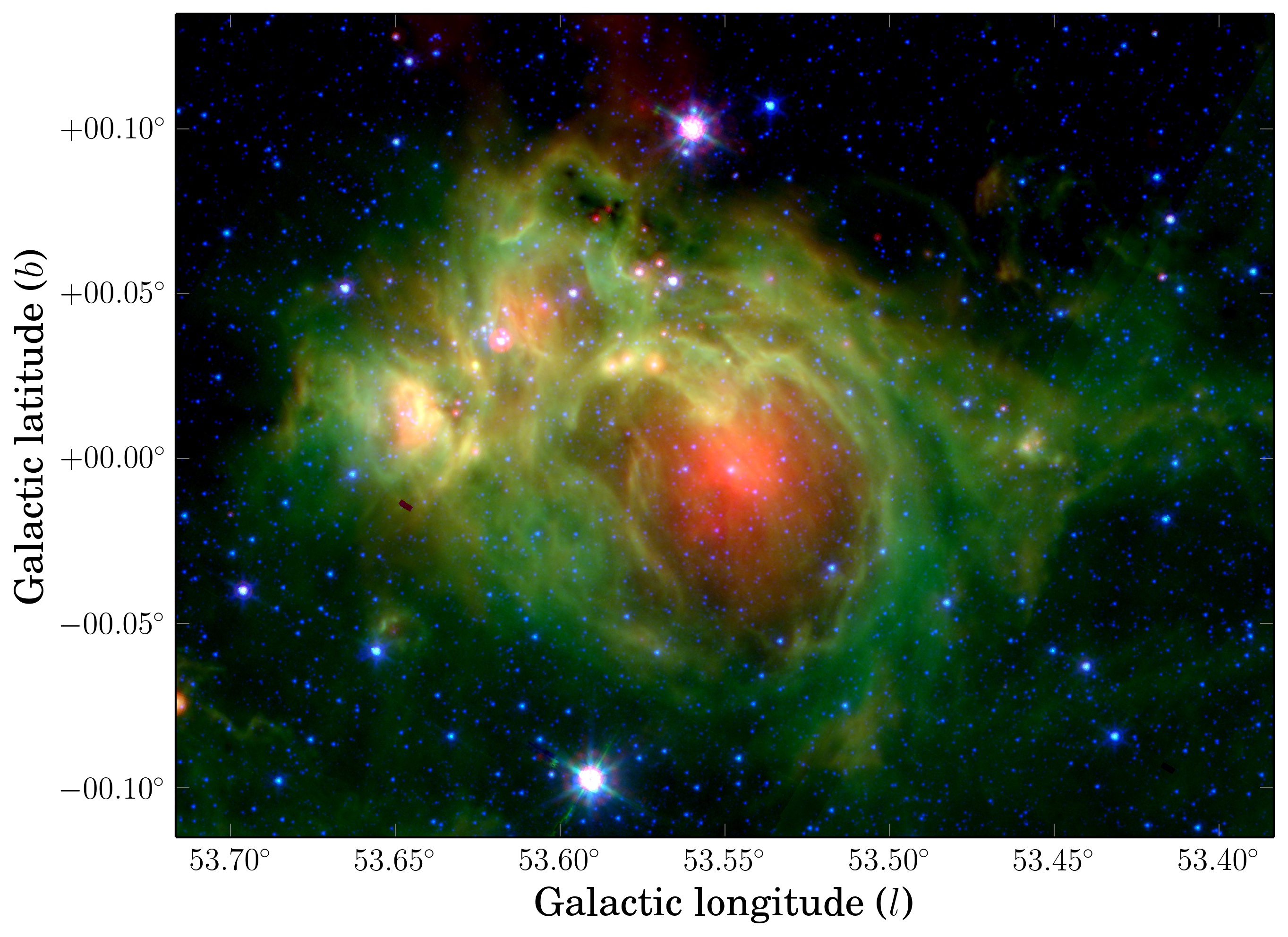} 
    \end{minipage}
    \caption{Star-forming region Sh 2-82 as seen in the mid-infrared
    by the Spitzer Space Telescope. The mosaic is composed of the 24~\micron\ (red), 8.0~\micron\ (green) and 4.5~\micron\ (blue) bands.
    The image reveals a bubble-shaped structure which surrounds the {\sc Hii} region that is seen in the IPHAS mosaic which spans the same region (Fig.~\ref{fig:mosaic_iphas}). 
    This structure has previously been labelled as N115 in the 
catalogue of \citet{Churchwell2006}, and could be a possible site of triggered star formation \citep{Thompson2012,Kendrew2012}.}
    \label{fig:mosaic_spitzer}
\end{figure*}

Fig.~\ref{fig:mosaic_iphas} shows a 20-by-15 arcmin
colour mosaic centred on Sh 2-82,
composed of our \ha\ (red channel),
$r$ (green channel),
and $i$ (blue channel) images.
The ionising star can be seen as the bright object
in the centre of the H{\sc ii} region,
which is surrounded by a faint reflection nebula
and several dark cloud filaments.
For comparison, Fig.~\ref{fig:mosaic_spitzer} shows
a mosaic of the same region 
as seen in the mid-infrared by the Spitzer Space Telescope \citep{Benjamin2003,Churchwell2009}.
The infrared image reveals an enclosing fuzzy bubble (appearing
green in Fig.~\ref{fig:mosaic_spitzer})
which is thought to originate from the
mid-infrared emission of Polycyclic Aromatic Hydrocarbons (PAHs)
-- i.e. warm dust --
which is frequently observed
at the interface between neutral regions of interstellar material
and the ionising radiation from early-type stars \citep{Churchwell2006}.
\cite{Yu2012}
recently noted that the warm dust
surrounding Sh~2-82 
appears to contain infrared-bright
Young Stellar Objects (YSOs).
Many of these young objects 
appear as red- and pink-coloured stars
in Fig.~\ref{fig:mosaic_spitzer},
located predominantly in the top-left part
of the bubble.

\begin{figure}
\begin{center}
  \includegraphics[width=\linewidth]{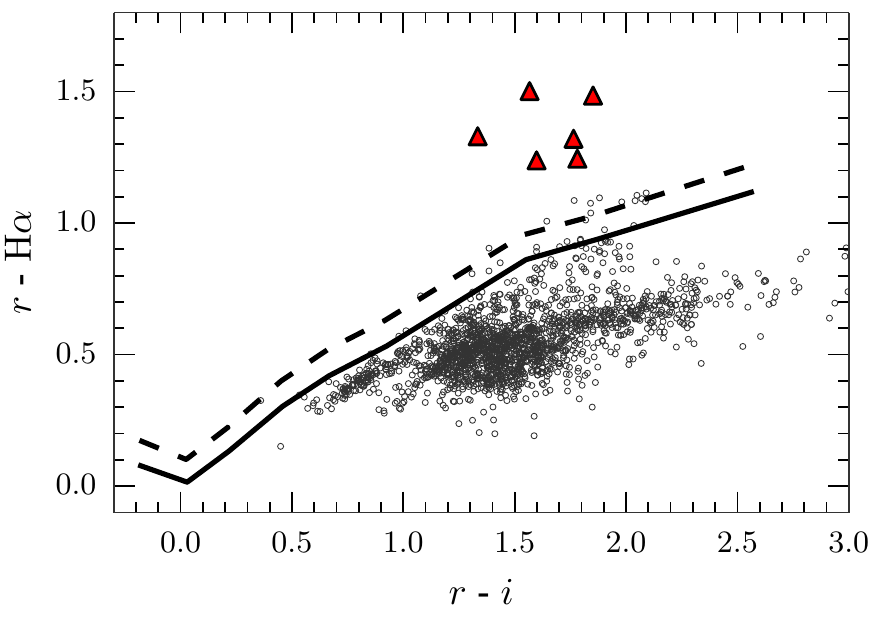}
    \caption{$(r-\ha,\ r-i)$ diagram for the rectangular region of 
    20-by-15 arcmin centred on the H{\sc ii} region Sh 2-82,
    which is the area shown in Fig.~\ref{fig:mosaic_iphas}.    
    The diagram shows all objects in the catalogue
    which have been flagged as \emph{a10} and are brighter
    than $r<20$ (grey circles).
    The unreddened main sequence is indicated by a solid line,
    while the main sequence for stars with an \ha\ emission line
    strength of $-10~\rm{\AA}$~EW is indicated by a dashed line
    \citep[both based on the colour simulations by][]{Barentsen2011a}.
    Red triangles indicate objects which have been identified as
    as likely \ha-emitters.}
    \label{fig:emitters}
\end{center}
\end{figure}

Fig.~\ref{fig:emitters} presents
the IPHAS colour-colour diagram for 
the 20-by-15 arcmin region shown in the mosaics.
Grey circles show all objects
which are brighter than $r<20$
and have been flagged as \emph{a10}
in IPHAS DR2.
The diagram also shows the unreddened main sequence (solid line)
and the expected position of unreddened main-sequence stars
with \ha\ in emission
at a strength of EW$=-10$~\AA\ (dashed line).
Six stars are found to lie above the 
dashed line at the level of $3\sigma$,
i.e. the distance between the objects and the dashed line
is larger than three times the uncertainty
in their $(r-\ha)$ colour.
These candidate \ha-emitters
are marked by red triangles in the colour-colour diagram,
and by yellow triangles in the image mosaic (Fig.~\ref{fig:mosaic_iphas}).
Their details are listed in Table~\ref{tbl:emitters}.

In previous work, we have shown that the majority of
\ha-emitters seen by IPHAS towards an H{\sc ii} region
are likely to be Classical T Tauri Stars \citep{Barentsen2011a}.
These are young objects
which are thought to show \ha\ in emission
due to the presence of hot, infalling gas
which is accreting onto the star
from a circumstellar disk.
This is likely to be the case for the candidate \ha-emitters
we discovered towards Sh 2-82 as well.
Two of our candidates, \#1 and \#4 in Table~\ref{tbl:emitters},
have previously been identified
as candidate YSOs by \cite{Robitaille2008}
and \cite{Yu2012}, respectively.
In these studies, the authors used Spitzer data
to find intrinsically red objects,
with SEDs consistent with the presence of a circumstellar disk.
Although the other four candidate emitters in our sample
have not previously appeared in the literature,
we note that all four are detected
in the Spitzer 8.0~\micron\ image at S/N$>$5.
They are likely to be YSOs exhibiting a mild infrared excess.
The recovery of \ha-emitters in Spitzer
data illustrates how IPHAS can complement infrared surveys.

\begin{table}
    \caption{Candidate \ha-emitters towards Sh 2-82.}
    \label{tbl:emitters}
    \footnotesize
    \resizebox{0.47\textwidth}{!}{
        \begin{tabular}{clccc}
        \toprule
        \# & Name [IPHAS2 ...] & $r$ & $i$ & \ha  \\
        \midrule
        1 & J192954.40+181026.1& $17.69\pm0.01$ & $16.12\pm0.01$ & $16.19\pm0.01$ \\
        2 & J193011.01+182051.2& $18.55\pm0.02$ & $16.95\pm0.02$ & $17.31\pm0.02$ \\
        3 & J193021.52+181954.5& $19.72\pm0.05$ & $17.94\pm0.03$ & $18.47\pm0.04$ \\
        4 & J193024.45+181938.3& $19.31\pm0.04$ & $17.55\pm0.02$ & $17.99\pm0.03$ \\
        5 & J193033.00+181609.3& $18.25\pm0.01$ & $16.91\pm0.01$ & $16.92\pm0.01$ \\
        6 & J193042.48+182317.4& $19.96\pm0.03$ & $18.11\pm0.03$ & $18.48\pm0.03$ \\
        \bottomrule
        \end{tabular}
    }
\end{table}

\subsection{IPHAS as a complement to infrared data}

\begin{figure}
    \includegraphics[width=\linewidth]{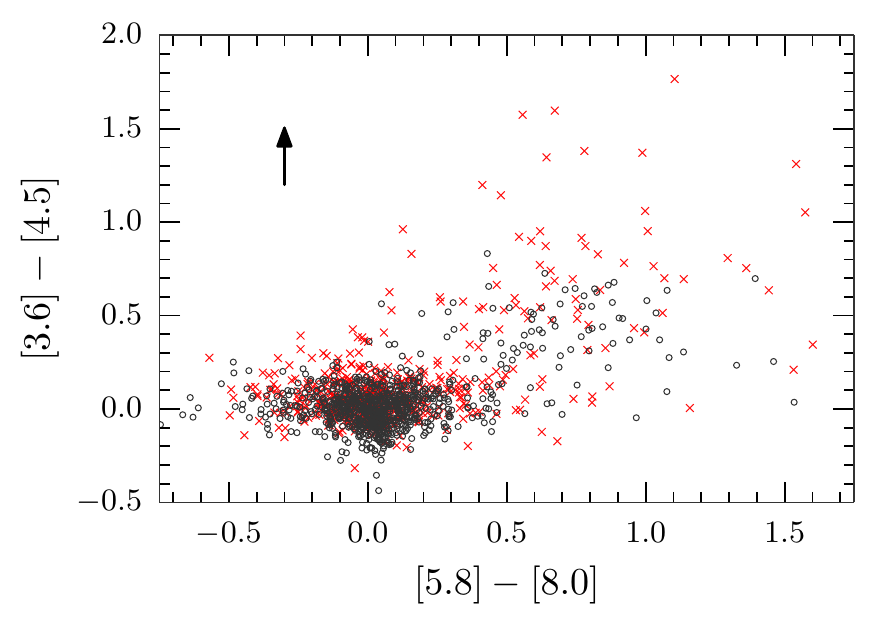} 
    \caption{Spitzer/GLIMPSE colour-colour diagram towards Sh~2-82,
    showing the objects for which a counterpart exists in IPHAS
    (845 objects, grey circles) 
    and those for which no counterpart was identified 
    (511 objects, red crosses).
    A cross-matching radius of 1 arcsec
    was used to identify counterparts.
    The arrow illustrates the reddening vector corresponding
    to $A_{Ks}=3$ following the reddening law due to \citet{Flaherty2007}.
    It is apparent that objects near the top of the diagram,
    where heavily reddened objects are expected to sit,
    tend to have no counterpart.
    In contrast, many of the objects on the right hand side of the diagram,
    where YSOs are expected to sit, do have counterparts.
    }
    \label{fig:spitzer-ccd}
\end{figure}

Towards Sh2-82,
the IPHAS and Spitzer/GLIMPSE catalogues
have 4,798 entries in common,
out of 10,739 and 11,321 entries in total, respectively.
Fig.~\ref{fig:spitzer-ccd} presents the Spitzer colour-colour diagram
of the region, showing the 1,356 objects that were detected 
in all four GLIMPSE bands.
We have indicated the 845 objects (62 per cent)
that have a counterpart in IPHAS (grey circles)
and the 511 objects that do not (red crosses).

We find that objects located in the upper half of the Spitzer diagram
are less likely to have an IPHAS counterpart,
which is where highly-reddened objects are expected to sit.
Moreover, amongst the objects with a counterpart,
we note that more have been detected in $i$ (843 objects)
than in $r$ (634) or \ha\ (600).
This is not surprising,
because Sh2-82 is located 
towards a high-extinction sightline 
near the inner plane,
where an optical survey will naturally be extinction-limited.
Nevertheless, the extinction map due to \citet{Sale2014}
has demonstrated that IPHAS can probe stars as far away 
as 4-5\,kpc towards this region.
The situation will be even more favourable
in the outer-plane, where dust optical depths are low.

We note that Sh 2-82 is one of a large population 
of poorly-studied star-forming regions
located at low Galactic latitudes,
which have only recently started to become revealed
by efforts to catalogue the wealth of `bubbles' detected
at mid-infrared wavelengths \citep{Churchwell2006,Simpson2012},
and by efforts to catalogue previously unknown clusters seen 
in the near-infrared \cite[e.g.][]{Bica2003}.
IPHAS data can offer a handle
on the extinction, distance
and stellar contents of many of such unexplored regions.

\subsection{$r - \ha$ excess as a quantitative measure of H$\alpha$ emission}

As well as discover emission line stars, IPHAS data can provide a
first estimate of the equivalent width of line emission.
How this is done has been discussed in previous works 
-- most notably by \citet{Drew2005}
and more recently by \cite{Barentsen2011a}.
Given that the narrowband filter used has a FWHM of 95~\AA ,
it is to be expected that the appearance
in a spectrum of line emission corresponding to an equivalent width (EW)
of 95~\AA\ would increase $r - \ha$ by 0.75 magnitudes
(i.e. the flux captured within the narrow band doubles).
Similarly, 10~\AA\ of line emission
should increase $r - H\alpha$ by 0.11~mag.

This simple reasoning breaks down
when the $H\alpha$ emission becomes so bright
that it also dominates the flux captured across the entire $r$ broadband.
It was noted by \cite{Drew2005}
that the very bright line limiting value 
of $r - \ha$ is $\sim$3.1 in the Vega system.
This fact, on its own, implies that any apparent detection
of an \ha\ emission line star without a corresponding $r$ 
detection is only real if the \ha\ magnitude is not brighter 
by more than 3.1~mag relative to the $r$ detection limit
(i.e. typically, credible H$\alpha$-only sources
must be fainter than $\sim$18th mag).  

In order to infer the emission EW from the $r - \ha$ colour
it is necessary to know the value this colour would have
in the case of no excess line emission.
In the {\em general} case this is not possible
without some prior knowledge of the star's reddening and spectral type
(i.e. the inferences drawn for an unreddened M dwarf or
a reddened F star would not be the same -- the absolute value of the 
equivalent width in the latter case would be greater because the 
reddened F star {\em without} emission would lie at lower $r -\ha$).
This is the situation faced in the study of IC 1396 by
\cite{Barentsen2011a} in which \ha\ EWs were inferred
relative to a reddened sequence of K-M stars
appropriate to this star-forming region.
A different choice of reddening would have resulted
in different EW estimates.

However it was found by \cite{Drew2005}
that for earlier type stars in which the spectrum
is not heavily modulated by line features or molecular bands
the estimation of net emission EW is more straightforward and unambiguous.  
It was demonstrated, for spectral slopes ranging
from the Rayleigh-Jeans limit to that appropriate to an early G star
and for a range of reddenings,
that it is possible to draw a unique set of constant equivalent-width 
lines in the $(r - \ha, r - i)$ diagram
that can be compared directly with the positions
of candidate emission line stars,
thereby predicting emission EW \citep[see fig.~6, ][]{Drew2005}.
The EW in this case should be understood
to be the measure of the emission observed
above the interpolated continuum
(i.e. no correction is present
for any assumed infilled photospheric line absorption).

\begin{figure}
\begin{center}
  \includegraphics[width=\linewidth]{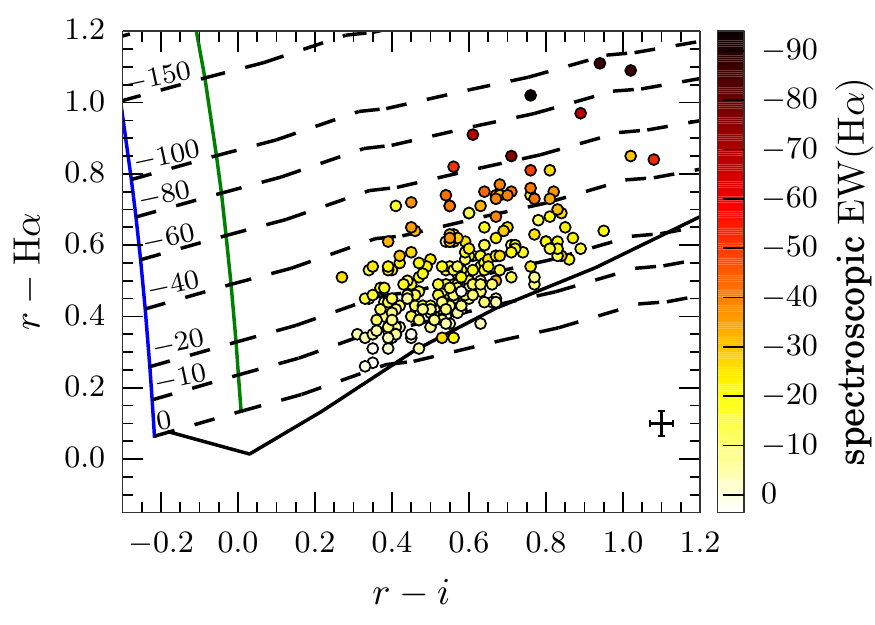}
    \caption{Photometric predictions of \ha\ equivalent width (EW)
    compared with spectroscopic measurements
    of a sample of classical Be stars (from Raddi et al 2014, submitted).
    The broken lines are predicted lines
    of constant net emission EW,
    while the analogous spectroscopic EW is represented
    by the colouring of the data points.
    The unreddened main sequence, which normally serves 
    as the upper bound to the main stellar locus,
    is drawn as a solid black line.
    Strongly reddened stars with not-so-strong line emission
    will fall below it
    and be hard to pick out without additional information.
    The two vertical lines drawn are 'curves of growth'
    showing how $r - H\alpha$ increases 
    -- and $r - i$ drops a little --
    as more and more line emission is added,
    raising the flux in \ha\ narrowband
    and to a lesser proportion the flux in the $r$ band.
    The blue line represents the trend
    for an unreddened Rayleigh-Jeans continuum
    (with superposed \ha\ emission),
    while the green line is the trend
    for the case of an unreddened A0 continuum.}
    \label{fig:raddi}
\end{center}
\end{figure}

In Fig.~\ref{fig:raddi} we show how well this works in reality:
the predicted lines of constant emission EW
in the $(r - \ha, r - i)$ plane
are compared with the positions of a sample
of over 200 classical Be stars in Perseus (from Raddi et al, 2014).
Stars of this type come with the advantage that secular variation,
whilst it certainly occurs,
is not as common as in T Tau stars.
The objects plotted are a bright selection ($r \lesssim 16.5$)
for which there are spectroscopic determinations
of \ha\ emission equivalent width.
A colour scale has been applied to the data points
that darkens as EW increases.
The gradation apparent in the shading of the datapoints
follows the trend set by the constant EW lines quite well.  
The experience has been that the EW deduced from
IPHAS photometry for individual objects is commonly within 10~\AA\
of subsequent spectroscopic measurement
\citep[see also fig.~5 of][]{Barentsen2011a}.
Evidently, the photometry is well-suited to statistical measurement
across large samples, while for individual objects a useful
approximation is delivered.

\section{Catalogue and image access}
\label{sec:dataaccess}

The catalogue will be made available through the
Vizier service (http://vizier.u-strasbg.fr),
where it can be queried using a web interface
and using Virtual Observatory (VO) protocols.
In addition, the catalogue can be downloaded
in its entirety from our website
as a collection of binary FITS tables,
each covering a $5\degr\times5\degr$ tile
of the footprint and comprising 50~gigabyte
in total (see www.iphas.org/dr2).

We do not recommend using the catalogue
to study extended objects which
are larger than the
aperture diameters specified in this work.
To enable the analysis of diffuse sources,
our website provides access to the entire set
of pipeline-processed imaging data and
associated meta data
(see www.iphas.org/images).
The image headers have been updated to include
a new keyword, called {\sc photzp},
which contains the re-calibrated zeropoint.
This keyword can be used to convert the number counts $DN$,
i.e. the pixel values in the images,
into Vega-based magnitudes $m$ using:
\begin{equation}
\begin{split}
   m  = \textsc{photzp} - 2.5 \log_{10}(DN).
\label{eqn:mag}
\end{split}
\end{equation}
The {\sc photzp} value has been computed such that it
absorbs the required corrections for atmospheric extinction,
gain variations, exposure time, and the re-calibration shift.
As these images still include moonlight and other sources
of non-astronomical background, they can only support flux measurements
that include a suitably-chosen local background subtraction.

To estimate absolute narrow-band \ha\ fluxes from the image data,
we note that the integrated in-band energy flux for Vega 
in the IPHAS \ha\ filter 
is $1.52 \times 10^{-7}$~erg\,cm$^{-2}$\,s$^{-1}$ 
at the top of the Earth's atmosphere,
which is the flux obtained by folding the 
CALSPEC SED with the filter transmission curve only
(the correction for atmosphere and detector quantum efficiency,
otherwise scales down the narrowband flux by 0.707).
This implies that the in-band flux corresponding to
zero magnitude is $1.56 \times 10^{-7}$ erg\,cm$^{-2}$\,s$^{-1}$,
when the \ha\ magnitude for Vega is set by convention to 0.03 \citep{Fukugita1996}.
These flux estimates are consistent with the most recent
version of Vega's SED presented by \citet{Bohlin2014}.

We warn that the image repository on our website
includes data that did not pass quality control
and has not been globally re-calibrated.
Such data are flagged in the on line meta data table,
which is available from our website,
and must be used with great caution.

In the spirit of reproducibility,
the source code that was used to generate
the catalogue is made available at
https://github.com/barentsen/iphas-dr2

\section{Concluding remarks and future work}
\label{sec:conclusions}

A new catalogue has been
derived from the INT/WFC Photometric \ha\ Survey
of the Northern Galactic Plane.
It is the first to offer comprehensive CCD photometry
of point sources across the Northern Galactic Plane at visible wavelengths,
taking in the Galactic latitude range $|b|<5\degr$
at longitudes $\ell=30\degr$ to 215\degr.
The new 99-column catalogue provides single-epoch photometry
across 92 per cent of the survey area,
and is the first quality-controlled and globally calibrated
catalogue to have been constructed from the imaging data.  This
now means that there is H$\alpha$ coverage, accessible online,
of the entire Galactic Plane -- given that the 
southern Plane is already available thanks to the UK
Schmidt H$\alpha$ Survey \citep[SHS,][]{Parker2005}, the last of the 
photographic surveys carried out by that telescope.

The observations included in this release
achieve a median seeing of 1.1 arcsec
and $5\sigma$-depths of $r=21.2\pm 0.5$, $i=20.0\pm 0.3$, and \ha$=20.3\pm 0.3$.
The global calibration and photometric repeatability
are found to be accurate at the level of $0.03$ mag (rms),
providing a significant improvement over the 
previous data release.
The source catalogue specifies the best-available
single-epoch astrometry and photometry
for 219~million unique sources.
To support its exploitation, we provide a list of recommended quality criteria
that will permit the selection of objects with accurate colours from 
the catalogue.  The closing demonstrations highlight the use of the 
survey's unique $(r-\ha,\ r-i)$ diagram for characterising stellar populations
and selecting emission-line objects.  More comprehensive applications of IPHAS
can be found in the works of \citet{Sale2014}, which applies DR2 to the problem
of 3D extinction mapping, and of \citet{Sabin2014}, where the results of a search
of the image database for new planetary nebulae is presented. 

The current plan is to work toward
one further major IPHAS source catalogue,
in which the remaining gaps in sky coverage
will have been eliminated
-- observations aimed at replacing data
not meeting the quality requirements are continuing.
We will also examine options to further improve
the global calibration,
perhaps tightening the accuracy to better than 2 per cent.
For example, we have in mind investigating
the use of the PanSTARRS photometric ladder \citep{Magnier2013}
as a reference set,
when it becomes available for the Galactic Plane, and we will 
explore improving source recovery in 
the most dense fields via the implementation of PSF fitting
in place of aperture photometry.  
Finally, the next catalogue will detail all the secondary detections
to aid time-domain studies.

The data-taking strategy developed for IPHAS
has since been reapplied to carry out 
a companion INT/WFC Galactic Plane survey called UVEX
in U, $g$, $r$, He {\sc i} \citep{Groot2009},
a survey of the Kepler field 
in $U$, $g$, $r$, $i$, \ha\ 
\citep{Greiss2012},
and a survey of the Southern Galactic Plane and Bulge
in $u$, $g$, $r$, $i$, \ha\ 
called VPHAS+ \citep{Drew2014}.  The last of these incorporates the
digital update of the SHS, offering all the advantages of calibrated
photometry across a little over half the SHS footprint.  
The work presented here
stands as a potential template for the catalogues that
remain to be generated from these sibling surveys.  
In prospect from them, whether they are mined separately or together,
are the means to ask seamless questions on the contents and structure of 
the most highly-populated components of the Milky Way.  

% woof woof ／人 ◕‿‿◕ 人＼ woof woof

\section*{Acknowledgments}

The INT is operated on the island of La Palma
by the Isaac Newton Group (ING)
in the Spanish Observatorio del Roque de los Muchachos
of the Instituto de Astrof\'\i sica de Canarias.
We are deeply indebted to Ovidiu Vaduvescu,
Javier M\'endez and the rest of the ING staff and students
for their ongoing support of the telescope.
All data were processed 
by the Cambridge Astronomical Survey Unit
at the Institute of Astronomy in Cambridge.
The catalogue presented in this work was assembled
at the Centre for Astrophysics Research, University of Hertfordshire, supported by a grant from the Science \& Technology Facilities Council
of the UK (STFC, ref ST/J001335/1).

Preparation of the catalogue was eased greatly
by a number of software packages,
including the {\sc postgresql} database software,
the {\sc topcat} and {\sc stilts} packages \citep{Taylor2005,Taylor2006},
and the Python modules
{\sc astropy} \citep{Astropy},
{\sc numpy} and {\sc scipy} \citep{Numpy},
{\sc matplotlib} \citep{Matplotlib},
{\sc ipython} \citep{IPython},
and {\sc aplpy}.
We also made use of the {\sc montage} software maintained by NASA/IPAC,
and the {\sc simbad}, {\sc vizier} and {\sc aladin} services
operated at CDS, Strasbourg, France \citep{Aladin}.

Our work made extensive use of
several complementary photometric surveys.
Our global calibration was aided
by the AAVSO Photometric All-Sky Survey (APASS),
funded by the Robert Martin Ayers Sciences Fund.
The calibration was tested against the
Sloan Digitized Sky Survey (SDSS),
funded by the Alfred P. Sloan Foundation,
the Participating Institutions,
the National Science Foundation,
the U.S. Department of Energy,
the National Aeronautics and Space Administration,
the Japanese Monbukagakusho, the Max Planck Society,
and the Higher Education Funding Council for England.
The astrometric pipeline reduction made
significant use of the Two Micron All Sky Survey (2MASS),
which is a joint project 
of the University of Massachusetts
and the Infrared Processing and Analysis Center/
California Institute of Technology,
funded by NASA and the NSF.
This work includes observations made
with the Spitzer Space Telescope,
which is operated by the Jet Propulsion Laboratory,
California Institute of Technology under a contract with NASA. 

GB, JED, SES and BTG acknowledge support from the Science \& Technology
Facilities Council of the United Kingdom
(grants: GB and JED ST/J001333/1, SES ST/K00106X/1, BTG ST/I001719/1). 
HJF and MM-S both acknowledge STFC postgraduate studentships.
JED would also like to convey her thanks
to the Physics Department of Imperial College London
that hosted this project from its inception to 2007 and supported
her via a sabbatical year in 2003-4.
JF is supported by the Spanish Plan Nacional de I+D+i and FEDER under contract AYA2010-18352.
BTG acknowledges funding from the
European Research Council under the European Union's Seventh Framework
Programme (FP/2007-2013) / ERC Grant Agreement n. 320964 (WDTracer).
PRG is supported by a Ram\'on y Cajal fellowship (RYC-2010-05762), and acknowledges support provided by the Spanish MINECO AYA2012--38700 grant.
NJW is in receipt of a Fellowship funded by 
the Royal Astronomical Society of the United Kingdom.

We are grateful to the anonymous referee for the insightful feedback,
which has helped us to improve the presentation of the data products.

\label{lastpage}

\footnotesize{
  \bibliographystyle{mn2e}
  \bibliography{iphas-dr2}
}

\newpage
\appendix

\newpage
\onecolumn
\section{Catalogue format}
\label{app:columns}

\small
\begin{longtable}{rlllp{10cm}}
\caption{\label{tab:columns} 
Definition of columns in the IPHAS DR2 source catalogue.
} \\
\hline
\# & Column & Type & Unit & Description \\
\hline
\endfirsthead

\multicolumn{3}{c}%
{{\bfseries \tablename\ \thetable{} -- continued}} \\
\hline
\# & Column & Type & Unit & Description \\
\hline
\endhead

\hline \hline
\endlastfoot
\input{tables/columns.tex}
\end{longtable}
\normalsize
\twocolumn

\end{document}

%% file: tables/columns.tex
1 & name & string &  & Position-based source name in the sexagesimal form: "JHHMMSS.ss+DDMMSS.s". You need to add the prefix "IPHAS2" followed by a whitespace to obtain the official name "IPHAS2 JHHMMSS.ss+DDMMSS.s" (where "J" indicates that the position is J2000 equatorial and "IPHAS2" indicates DR2). \\
2 & ra & double & degrees & J2000 Right Ascension with respect to the 2MASS PSC reference frame, which is consistent with ICRS to within 0.1 arcsec. The coordinate given is obtained from the astrometric measurement in the r-band exposure. If the source is undetected in r, then the i or H$\alpha$-band coordinate is given. \\
3 & dec & double & degrees & J2000 Declination. See comments above. \\
4 & sourceID & string &  & Unique identification number of the detection. Identical to rDetectionID if the source was detected in the r-band. Identical to iDetectionID or haDetectionID otherwise. \\
5 & posErr & float & arcsec & Astrometric root mean square (RMS) residual measured against 2MASS across the CCD in which the source is detected. Be aware that the astrometric error for a source near the corner of a CCD may be significantly larger than the RMS statistic. \\
6 & l & double & degrees & Galactic longitude $\ell$ converted from ra/dec (IAU 1958 system). \\
7 & b & double & degrees & Galactic latitude $b$ converted from ra/dec (IAU 1958 system). \\
8 & mergedClass & short &  & Image classification flag based on all bands: 1=galaxy, 0=noise, -1=star, -2=probableStar, -3=probableGalaxy, -9=saturated. Computed using the UKIDSS scheme. \\
9 & mergedClassStat & float &  & Merged N(0,1) stellarness-of-profile statistic. Computed using the UKIDSS scheme. \\
10 & pStar & float &  & Probability that the source is a point source (value between 0 and 1). \\
11 & pGalaxy & float &  & Probability that the source is an extended object, such as a galaxy, or a close blend of two point sources (value between 0 and 1). \\
12 & pNoise & float &  & Probability that the source is noise, e.g. a cosmic ray (value between 0 and 1). \\
13 & rmi & float & mag & (r - i) colour, formed by subtracting columns r and i.  To obtain the uncertainty, take the root of the sum of the squares of columns rErr and iErr. \\
14 & rmha & float & mag & (r - Halpha) colour, formed by subtracting columns r and ha. See comments above. \\
15 & r & float & mag & Default r-band magnitude using the 2.3 arcsec diameter aperture. Calibrated in the Vega system. \\
16 & rErr & float & mag & Uncertainty for r. Does not include systematic errors. \\
17 & rPeakMag & float & mag & Alternative r-band magnitude derived from the peak pixel height (i.e. a 0.3x0.3 arcsec square aperture). Calibrated in the Vega system. \\
18 & rPeakMagErr & float & mag & Uncertainty in rPeakMag. Does not include systematics. \\
19 & rAperMag1 & float & mag & Alternative r-band magnitude using the 1.2 arcsec diameter aperture. Calibrated in the Vega system. \\
20 & rAperMag1err & float & mag & Uncertainty in rAperMag1. Does not include systematics. \\
21 & rAperMag3 & float & mag & Alternative r-band magnitude using the 3.3 arcsec diameter aperture. Calibrated in the Vega system. \\
22 & rAperMag3err & float & mag & Uncertainty in rAperMag3. Does not include systematics. \\
23 & rGauSig & float & pixels & RMS of axes of ellipse fit in r. \\
24 & rEll & float &  & Ellipticity in the r-band. \\
25 & rPA & float & degrees & Position angle in the r-band. \\
26 & rClass & short &  & Discrete image classification flag: 1=galaxy, 0=noise, -1=star, -2=probableStar, -3=probableGalaxy, -9=saturated. \\
27 & rClassStat & float &  & N(0,1) stellarness-of-profile statistic. \\
28 & rDeblend & boolean &  & True if the source is blended with a nearby neighbour in the r-band. Although a deblending procedure is applied when measuring the photometry, the result may be unreliable (colours should not be trusted in particular). \\
29 & rSaturated & boolean &  & True if the source is too bright to make an accurate measurement in the r-band (e.g. peak pixel $>$ 55000 counts). The photometry is likely affected by systematic errors. \\
30 & rMJD & double & days & Modified Julian Date at the start of the r-band exposure. \\
31 & rSeeing & float & arcsec & Average Full Width at Half Maximum (FWHM) of stars in the same CCD frame. \\
32 & rDetectionID & string &  & Unique identifier of the r-band detection in the format "$\#$run-$\#$ccd-$\#$number", i.e. composed of the INT telescope run number, the CCD number and a sequential source detection number. \\
33 & rX & float & pixels & Pixel coordinate of the source in the r-band exposure, in the coordinate system of the CCD. \\
34 & rY & float & pixels & Pixel coordinate of the source in the r-band exposure, in the coordinate system of the CCD. \\
35 & i & float & mag & Default i-band magnitude using the 2.3 arcsec diameter aperture. Calibrated in the Vega system. \\
36 & iErr & float & mag & Uncertainty for i. Does not include systematic errors. \\
37 & iPeakMag & float & mag & Alternative i-band magnitude derived from the peak pixel height (i.e. a 0.3x0.3 arcsec square aperture). Calibrated in the Vega system. \\
38 & iPeakMagErr & float & mag & Uncertainty in iPeakMag. Does not include systematics. \\
39 & iAperMag1 & float & mag & Alternative i-band magnitude using the 1.2 arcsec diameter aperture. Calibrated in the Vega system. \\
40 & iAperMag1err & float & mag & Uncertainty in iAperMag1. Does not include systematics. \\
41 & iAperMag3 & float & mag & Alternative i-band magnitude using the 3.3 arcsec diameter aperture. Calibrated in the Vega system. \\
42 & iAperMag3err & float & mag & Uncertainty in iAperMag3. Does not include systematics. \\
43 & iGauSig & float & pixels & RMS of axes of ellipse fit. \\
44 & iEll & float &  & Ellipticity. \\
45 & iPA & float & degrees & Position angle. \\
46 & iClass & short &  & Discrete image classification flag: 1=galaxy, 0=noise, -1=star, -2=probableStar, -3=probableGalaxy, -9=saturated. \\
47 & iClassStat & float &  & N(0,1) stellarness-of-profile statistic. \\
48 & iDeblend & boolean &  & True if the source is blended with a nearby neighbour in the i-band. See comments for rDeblend above. \\
49 & iSaturated & boolean &  & True if the source is too bright to make an accurate measurement in the i-band. See comments for rSaturated above. \\
50 & iMJD & double & days & Modified Julian Date at the start of the single-band exposure. \\
51 & iSeeing & float & arcsec & Average Full Width at Half Maximum (FWHM) of stars in the same CCD frame. \\
52 & iDetectionID & string &  & Unique identifier of the i-band detection in the format "$\#$run-$\#$ccd-$\#$number", i.e. composed of the INT telescope run number, the CCD number and a sequential source detection number. \\
53 & iX & float & pixels & Pixel coordinate of the source, in the coordinate system of the CCD. \\
54 & iY & float & pixels & Pixel coordinate of the source, in the coordinate system of the CCD. \\
55 & iXi & float & arcsec & Position offset of the i-band detection relative to the ra column. The original i-band coordinates can be obtained by computing (ra+iXi/3600, dec+iEta/3600). \\
56 & iEta & float & arcsec & Position offset of the i-band detection relative to the dec column. See comments above. \\
57 & ha & float & mag & Default H-alpha magnitude using the 2.3 arcsec aperture. Calibrated in the Vega system. \\
58 & haErr & float & mag & Uncertainty for ha. Does not include systematic errors. \\
59 & haPeakMag & float & mag & Alternative H-alpha magnitude derived from the peak pixel height (i.e. a 0.3x0.3 arcsec square aperture). Calibrated in the Vega system. \\
60 & haPeakMagErr & float & mag & Uncertainty in haPeakMag. Does not include systematics. \\
61 & haAperMag1 & float & mag & Alternative H-alpha magnitude using the 1.2 arcsec diameter aperture. Calibrated in the Vega system. \\
62 & haAperMag1err & float & mag & Uncertainty in haAperMag1. Does not include systematics. \\
63 & haAperMag3 & float & mag & Alternative H-alpha magnitude using the 3.3 arcsec diameter aperture. Calibrated in the Vega system. \\
64 & haAperMag3err & float & mag & Uncertainty in haAperMag3. Does not include systematics. \\
65 & haGauSig & float & pixels & RMS of axes of ellipse fit. \\
66 & haEll & float &  & Ellipticity. \\
67 & haPA & float & degrees & Position angle. \\
68 & haClass & short &  & Discrete image classification flag: 1=galaxy, 0=noise, -1=star, -2=probableStar, -3=probableGalaxy, -9=saturated. \\
69 & haClassStat & float &  & N(0,1) stellarness-of-profile statistic. \\
70 & haDeblend & boolean &  & True if the source is blended with a nearby neighbour in H-alpha. See comments for rDeblend above. \\
71 & haSaturated & boolean &  & True if the source is too bright to make an accurate measurement in H-alpha. See comments for rSaturated above. \\
72 & haMJD & double & days & Modified Julian Date at the start of the single-band exposure. \\
73 & haSeeing & float & arcsec & Average Full Width at Half Maximum (FWHM) of stars in the same CCD frame. \\
74 & haDetectionID & string &  & Unique identifier of the H-alpha detection in the format "$\#$run-$\#$ccd-$\#$number", i.e. composed of the INT telescope run number, the CCD number and a sequential source detection number. \\
75 & haX & float & pixels & Pixel coordinate of the source, in the coordinate system of the CCD. \\
76 & haY & float & pixels & Pixel coordinate of the source, in the coordinate system of the CCD. \\
77 & haXi & float & arcsec & Position offset of the H-alpha detection relative to the ra column. The original Ha-band coordinates can be obtained by computing (ra+haXi/3600, dec+haEta/3600). \\
78 & haEta & float & arcsec & Position offset of the H-alpha relative to the ra column. See comments above. \\
79 & brightNeighb & boolean &  & True if a very bright star is nearby (defined as brighter than V$<$4 within 10 arcmin, or brighter than V$<$7 within 5 arcmin).  Such very bright stars cause scattered light and diffraction spikes, which may add systematic errors to the photometry or even trigger spurious detections. \\
80 & deblend & boolean &  & True if the source is blended with a nearby neighbour in one or more bands. Although a deblending procedure is applied when measuring the photometry, the result may be inaccurate and the colours should not be trusted. \\
81 & saturated & boolean &  & True if the source is saturated in one or more bands. The photometry of saturated stars is affected by systematic errors. \\
82 & nBands & short &  & Number of bands in which the source is detected (equals 1, 2 or 3). \\
83 & a10 & boolean &  & True if the source is detected at S/N\,$>$\,10 in all bands without being saturated, and if the photometric measurements are consistent across different aperture diameters. Algebraic condition: (rErr\,$<$\,0.1 \& iErr\,$<$\,0.1 \& haErr\,$<$\,0.1 \& NOT saturated \& (abs(r-rAperMag1)\,$<$\,3*hypot(rErr,rAperMag1Err)+0.03) \& (abs(i-iAperMag1)\,$<$\,3*hypot(iErr,iAperMag1Err)+0.03) \& (abs(ha-haAperMag1)\,$<$\,3*hypot(haErr,haAperMag1Err)+0.03). \\
84 & a10point & boolean &  & True if both the a10 quality criteria above are satisfied, and if the object looks like a single, unconfused point source. Algebraic condition: a10 \& pStar\,$>$\,0.9 \& NOT deblend \& NOT brightNeighb. \\
85 & fieldID & string &  & Survey field identifier (e.g. 0001\_aug2003). \\
86 & fieldGrade & string &  & Internal quality control score of the field. One of A, B, C or D. \\
87 & night & integer &  & Night of the observation (YYYYMMDD). Refers to the UT date at the start of the night. \\
88 & seeing & float & arcsec & Maximum value of rSeeing, iSeeing, or haSeeing. \\
89 & ccd & short &  & CCD-chip number on the Wide Field Camera (WFC) of the Isaac Newton Telescope (INT). 1, 2, 3 or 4. \\
90 & nObs & short &  & Number of repeat observations of this source in the survey. A value larger than 1 indicates that the source is unlikely to be spurious. \\
91 & sourceID2 & string &  & SourceID of the alternative detection of the object in the partner exposure. \\
92 & fieldID2 & string &  & FieldID of the partner detection (e.g. 0001o\_aug2003). \\
93 & r2 & float & mag & r-band magnitude in the dithered partner field, i.e. the dithered repeat measurement obtained within 10 minutes (if available). \\
94 & rErr2 & float & mag & Uncertainty for r2. \\
95 & i2 & float & mag & i-band magnitude in the dithered partner field, i.e. the dithered repeat measurement obtained within 10 minutes (if available). \\
96 & iErr2 & float & mag & Uncertainty for i2. \\
97 & ha2 & float & mag & H-alpha magnitude in the dithered partner field, i.e. the dithered repeat measurement obtained within 10 minutes (if available). \\
98 & haErr2 & float & mag & Uncertainty for ha2. \\
99 & errBits2 & integer &  & Error bitmask for the partner detection. Used to flag a bright neighbour (1), source blending (2), saturation (8), vignetting (64), truncation (128) and bad pixels (32768).  Be careful if errBits2\,$>$\,0. \\